\documentclass[
amsmath,
amssymb,
aps,
prl,
reprint,
floatfix,
superscriptaddress,
longbibliography
]{revtex4-2}

\usepackage[T1]{fontenc}
\usepackage[utf8]{inputenc}
\usepackage{graphicx}
\usepackage{xcolor}
\definecolor{linkscolor}{RGB}{10,55,130}
\usepackage{bm}
\usepackage{braket}
\usepackage{times}
\usepackage{physics}
\usepackage{units}
\usepackage{upgreek}
\usepackage[percent]{overpic}

\usepackage[
pdftex,
colorlinks=true,
linkcolor=linkscolor,
citecolor=linkscolor,
urlcolor=linkscolor,
breaklinks=true
]{hyperref}

\newcommand{\bea}{\begin{eqnarray}}
\newcommand{\eea}{\end{eqnarray}}
\newcommand{\beq}{\begin{equation}}
\newcommand{\eeq}{\end{equation}}
\newcommand{\benu}{\begin{enumerate}}
\newcommand{\enu}{\end{enumerate}}

\newcommand{\al}{\alpha}
\newcommand{\be}{\beta}

\newcommand{\om}{\omega}
\newcommand{\Om}{\Omega}
\newcommand{\ep}{\epsilon}

\newcommand{\si}{\sigma}

\newcommand{\ham}{\mathcal{H}}

\newcommand{\cda}{c^{\dagger}}
\newcommand{\bk}{{\bf k}}
\newcommand{\bq}{{\bf q}}

\begin{document}

\title{
Electron-Mediated Phonon Polaritons in Quantum Cavities}
\date{\today}
\author{Mattia Udina}
\email[]{mattia.udina@mpsd.mpg.de}
\affiliation{Laboratoire Mat\'{e}riaux et Ph\'{e}nom\`{e}nes Quantiques,
Universit\'{e} Paris Cit\'{e}, CNRS, 75205 Paris, France}
\affiliation{Institut de Physique et Chimie des Mat\'{e}riaux de Strasbourg (UMR 7504),
Universit\'{e} de Strasbourg and CNRS, Strasbourg, 67200, France}

\author{David Hagenm\"{u}ller}
\email[]{david.hagenmuller@ipcms.unistra.fr}
\affiliation{Institut de Physique et Chimie des Mat\'{e}riaux de Strasbourg (UMR 7504),
Universit\'{e} de Strasbourg and CNRS, Strasbourg, 67200, France}

\author{Indranil Paul}
\email[]{indranil.paul@u-paris.fr}
\affiliation{Laboratoire Mat\'{e}riaux et Ph\'{e}nom\`{e}nes Quantiques,
Universit\'{e} Paris Cit\'{e}, CNRS, 75205 Paris, France}

\begin{abstract}
Phonon--photon coupling in quantum materials is commonly treated as an
effective direct interaction, neglecting the crucial role played by electrons. Here we develop a gauge-invariant theory of cavity phonon polaritons that captures electronic screening, dissipation, and hybridization. We show that screening can suppress polariton formation in metals, severely constraining cavity control of phonon-mediated superconductivity. By contrast, interband electronic resonances in time-reversal-symmetric gapped systems can strongly enhance phonon--photon coupling, enabling robust polariton formation down to the monolayer limit despite electronic absorption. We illustrate these effects in NbSe$_2$, hBN, and gapped graphene. Finally, we show that gauge invariance enforces photon decoupling from a critical soft phonon, establishing a general no-go mechanism for photon condensation.
\end{abstract}

\maketitle

\emph{Introduction.}---Advances in electromagnetic field confinement have opened new routes to control condensed-matter phenomena, giving rise to cavity quantum materials~\cite{Bloch2022,Schlawin2022,Hubener2024,Bretscher2026FluctuationEngineering}. Two-dimensional (2D) materials are particularly promising owing to their rich and tunable electronic properties~\cite{Liu2015,Flatten2016,Liu2016,Zheng2017,Chen2017,Kleemann2017,Manzeli2017,Chervy2018,Stuhrenberg2018,Wilson2021,Dirnberger2023,Nguyen2023,Nuckolls2024,Liu2025CavityMediated}. Recent experiments have demonstrated strong light--matter interactions with deeply subwavelength modes supported by external photonic structures~\cite{Helmrich2026,Jochl2026}, plasmonic self-cavities~\cite{Kipp2025CavityElectrodynamics}, and intrinsic polaritonic modes~\cite{Koppens2011GraphenePlasmonics,Dai2014,Caldwell2014,Brar2014,Li2015,Autore2018,Sheinfux2024,Wang2024,Riolo2025Tuning,Andolina2026}. A major research direction is cavity control of superconductivity, with recent experimental~\cite{Keren2026CavityAltered,Xu2026VacuumDressed,Montanaro2026CavityEnhanced,Zhang2026CavityEnhanced,Wang2026VacuumEnhanced} and theoretical~\cite{Schlawin2019,Curtis2019,Lu2024CavityEnhanced,AndolinaAmperean2024,Kozin2025,Jiang2026NoGo} advances. For conventional superconductors, an early and direct approach involves coupling cavity photons to lattice vibrations mediating Cooper pairing~\cite{Sentef2018CavityQED,Thomas2025Exploring}.

This approach relies on phonon--photon hybridization, whose microscopic origin in quantum materials remains poorly understood. Most theories assume a direct effective coupling between infrared-active phonons and cavity photons~\cite{Sentef2018CavityQED,Thomas2025Exploring}. Microscopically, however, this coupling is strongly influenced by electrons, which also govern electromagnetic screening and dissipation. Their interplay is particularly important in conductors, where electronic screening can suppress light--matter hybridization, as recently observed across a metal--insulator transition~\cite{Jarc2024Multimode}. A consistent description therefore requires a gauge-invariant treatment of electron--photon coupling~\cite{Cottet2015ElectronPhoton,PhysRevA.98.053819,Andolina2019,Li2020,Dmytruk2021} that accounts for the material's electronic properties.

Here we develop a gauge-invariant microscopic theory of cavity phonon polaritons that explicitly incorporates electronic degrees of freedom. Integrating out the electrons yields two key response functions: the current--current susceptibility $\Pi_{jj}(\omega)$, governing screening and dissipation, and the mixed current--phonon susceptibility $\Pi_{j\Gamma}(\omega)$, determining
the electron-mediated phonon–photon coupling and generalizing the static phonon Born effective charge~\cite{stengel22,binci21}. We identify distinct regimes in metals, insulators, and semiconductors. In metals, the Drude response can prevent polariton formation, challenging recent proposals for cavity coupling to phonons in conventional superconductors~\cite{Sentef2018CavityQED,Thomas2025Exploring}. In insulators, the absence of low-energy electronic excitations suppresses screening and dissipation, allowing robust phonon--photon hybridization. Strikingly, we find that when the phonon is resonant with an interband transition in semiconductors, electronic excitations can instead promote polariton formation. Near this resonance, $\Pi_{j\Gamma}(\omega)$ is strongly enhanced and acquires sizable real and imaginary parts, enabling phonon--photon hybridization even in the monolayer limit despite electronic absorption. More generally, gauge invariance enforces $\Pi_{j\Gamma}(\omega\to0)=0$, decoupling cavity photons from critical soft phonons, e.g., near ferroelectric transitions~\cite{PhysRevX.10.041027} and extending no-go theorems for photon condensation to electron--phonon--photon systems. We demonstrate these effects using a tight-binding model capturing the main features of NbSe$_2$, hBN, and gapped graphene. Our results identify electronic resonance engineering as a route to enhance polariton formation through electron-mediated phonon--photon coupling.

\emph{Model and formalism}---We consider a 2D quantum material embedded in a Fabry-Perot cavity with photon confinement along the $z$-direction. The coupled light--matter system is described by the Hamiltonian $\ham = \ham_{\rm m} + \ham_{\rm l} + \ham_{\rm lm}$. The matter part is decomposed into three contributions, $\ham_{\rm m} = \ham_{\rm e} + \ham_{\rm p} + \ham_{\rm ep}$. The electrons are described by $\ham_{\rm e} = \sum_{\bk, \mu \nu, \si}  \ep_{\bk}^{\mu \nu} \cda_{\bk, \mu, \si} c^{\phantom{\dagger}}_{\bk, \nu, \si}$, where $\ep_{\bk}^{\mu \nu}$ is the electronic dispersion, $(\mu, \nu)$ and $\si$ are orbital and spin indices, respectively. In any relevant material, the pair of zone-centre optical phonons that couple linearly to in-plane electric fields $(E_x, E_y)$ transform as $(x, y)$ under the point group symmetries. These phonon branches are described by $\ham_{\rm p} = \sum_{\bq,\al} \hbar \om_{{\rm p},\bq} b^{\dagger}_{\bq,\al} b^{\phantom{\dagger}}_{\bq,\al}$, where $\om_{{\rm p},\bq}$ is the phonon dispersion and $\al=x,y$ labels the two phonon polarizations, which we assume to be degenerate for simplicity. In the above, $\bk$ and $\bq$ denote in-plane wavevectors. The electron--phonon interaction is given by
\begin{equation}
\ham_{\rm ep}=\frac{\lambda}{\sqrt{N}} \sum_{\mathbf{q}, \al}(b^{\phantom{\dagger}}_{\mathbf{q}, \al}
+b^{\dagger}_{\mathbf{-q}, \al})\hat \Gamma_{\mathbf{q}, \al},
\label{Hep}
\end{equation}
where $\lambda$ is the coupling energy, $N$ the number of lattice sites and $\hat \Gamma_{\mathbf{q}, \al} \equiv \sum_{\mathbf{k}\mu\nu\sigma}f^{\mu\nu}_{\mathbf{k,q}, \al}c_{\mathbf{k+q},\mu\sigma}^{\dagger}c^{\phantom{\dagger}}_{\mathbf{k},\nu\sigma}$. The form factor $f^{\mu\nu}_{\mathbf{k},\mathbf{q},\al}$ encodes the symmetry of the phonon displacement field $\hat{Q}_{\mathbf{q},\al} \equiv b^{\phantom{\dagger}}_{\mathbf{q},\al}+b_{-\mathbf{q},\al}^{\dagger}$. For example, in centrosymmetric systems, only polar phonons that are odd under inversion can couple to the electric field, implying $f_{-\mathbf{k},-\mathbf{q},\al}^{\mu\nu}=-f_{\mathbf{k},\mathbf{q},\al}^{\mu\nu}$. The cavity photon Hamiltonian reads $\ham_l=\sum_{\mathbf{q},\be}\hbar\omega_{\rm c,\mathbf{q}}
a^{\dagger}_{\mathbf{q},\be}a^{\phantom{\dagger}}_{\mathbf{q},\be}$, where $\omega_{\rm c,\mathbf{q}}=c\sqrt{(\pi/L)^2+q^2}$ is the dispersion of the fundamental cavity mode, with $q=\vert\mathbf{q}\vert$, $c$ the speed of light, $L$ the mirror separation, and $\be=1,2$ labeling the two transverse photon polarizations. In the dipole gauge, the light--matter interaction, comprising both the linear dipole coupling and the polarization self-energy, reads
\begin{equation}
\begin{aligned}
H_{\rm lm}=&  \sum_{\mathbf{q}, \be}\left [ \frac{U_{\mathbf{q}}}{\sqrt{N}}\hat D_{\mathbf{q},\be}
\mathbf{u}_{\bq, \be} \cdot \mathbf{\hat P_q}+ \frac{U_{\mathbf{q}}^2}{ N \hbar \omega_{{\rm c},\mathbf{q}}} 
\left ( \bf{u}_{\bq, \be} \cdot \mathbf{\hat P_q} \right )^2 \right],
\end{aligned}
\label{Hlm}
\end{equation}
where $\hat D_{\mathbf{q}, \be}\equiv \mathrm{i}(a^{\phantom{\dagger}}_{\mathbf{q}, \be}-a_{\mathbf{-q}, \be}^{\dagger})$ are the photon displacement fields, $U_{\mathbf{q}}=e\sqrt{\hbar \omega_{{\rm c},\mathbf{q}}/(2\epsilon_0L)}$ is the electron-photon coupling strength, $\bf{u}_{\bq, \be}$ is the photon polarization vector, and $\hat{\mathbf{P}}_{\mathbf{q}}$ is the Fourier transform of the dimensionless electronic polarization operator. Hereafter, we set $\hbar=1$. Within the Peierls approximation, which preserves gauge invariance~\cite{Li2020,Dmytruk2021}, $\hat{\mathbf{P}}_{\mathbf{R}}\equiv \sum_{\mu, \sigma} \mathbf{R}c_{\mathbf{R}\mu,\sigma}^{\dagger}c^{\phantom{\dagger}}_{\mathbf{R}\mu,\sigma}$. We neglect an explicit ionic contribution to the polarization \cite{fachin24}, which does not qualitatively affect our conclusions.

After integrating out the electrons, the coupled photon--phonon system is described by the effective action
\begin{equation}
S[D,Q]=\frac{1}{2} \sum_{\eta, \al, \be}\left [D_{-\eta, \al},Q_{-\eta, \be} \right] \hat M_{\al \be}(\eta) \begin{bmatrix} D_{\eta, \al} \\ Q_{\eta, \be} 
\end{bmatrix},
\end{equation}
where $\eta=(\omega,\mathbf{q})$. We evaluate the kernel $\hat M_{\al \be}(\eta)$ using diagrammatic techniques, with details provided in the Supplementary Material (SM)~\cite{Note1}. We summarize the main steps below. First, the photon mediated effective electron-electron interaction is given by 
$\widetilde W_{\mathbf{q}}(\om) \equiv (W_{\mathbf{q}, \om}+2/\omega_{\rm c,\mathbf{q}})$, 
with $W_{\mathbf{q}, \om} =2\omega_{\rm c,\mathbf{q}}/(\om^2-\omega_{\rm c,\mathbf{q}}^2)$ 
the bare photon propagator. Here, the two contributions are from the linear dipole coupling and the
polarization self-energy in Eq.~\eqref{Hlm}, respectively. Since $\widetilde W_{\mathbf{q}}(\omega)\propto \omega^2$, the light-matter coupling vanishes in the static limit $\om \rightarrow 0$. Later, we show that this
property forbids superradiance, i.e., photon condensation. Second, the form of $\widetilde W_{\mathbf{q}}(\om)$ guarantees that, in the evaluation of $\hat M_{\al \be}(\eta)$, the polarization operator always appears in the combination $\om \hat{\mathbf{P}}_{\mathbf{q}}$. Since $\om \hat{\mathbf{P}}_{\mathbf{q}} \propto \mathrm{i}[\ham,\hat{\mathbf{P}}_{\mathbf{q}}] \propto \mathbf{\hat j}_{\mathbf{q}}$, where $\mathbf{\hat j}_{\mathbf{q}}= \sum_{\bk, \mu \nu, \si} \nabla_{\bk} \ep_{\bk}^{\mu \nu} \cda_{\bk +\bq/2, \mu, \si} c^{\phantom{\dagger}}_{\bk - \bq/2, \nu, \si}$ is the current operator, we find that, even in the dipole gauge, the effective light--matter interaction involves the current matrix elements. This is conceptually satisfying, since the operator $\hat{\mathbf{P}}$ is ill defined in the thermodynamic limit \cite{resta}, whereas $\mathbf{\hat j}$ remains well defined. Third, in the terahertz-to-mid-infrared range of interest, photon momenta satisfy $qa_0\ll 1$, where $a_0$ is the lattice spacing, allowing the electronic susceptibilities to be evaluated in the uniform limit $q\to 0$. Furthermore, for $q\ll\pi/L$, the cavity has cylindrical symmetry, and the two transverse photon polarizations can be identified with the in-plane phonon polarizations. We adopt this approximation throughout, neglecting finite-$qL/\pi$ corrections at larger momenta. The kernel is then diagonal in polarization space, $\hat M_{\al\be}\propto\delta_{\al\be}$, yielding
\begin{equation}
\hat M_{\al \al}(\eta)=
\begin{pmatrix}
 W_{\mathbf{q}, \om}^{-1}-g_{\mathbf{q}}^2 \Pi_{j_{\al}j_{\al}}(\om) & 
\mathrm{i}g_{\mathbf{q}}\lambda \Pi_{j_{\al}\Gamma_{\al}}(\om) \\
-\mathrm{i}g_{\mathbf{q}}\lambda \Pi_{j_{\al} \Gamma_{\al}}(-\om) & B_{\bq, \om}^{-1}
-\lambda^2 \Pi_{\Gamma_{\al}\Gamma_{\al}}(\om)
\end{pmatrix},
\label{2x2}
\end{equation} 
with $g_{\mathbf{q}}= U_{\mathbf{q}} / \omega_{{\rm c},\mathbf{q}}$, 
and $B_{\bq, \om}=2\om_{{\rm p}, \bq}/(\om^2-\omega_{{\rm p}, \bq}^2)$ the bare phonon propagator.

In the above, the phonon--phonon correlator $\Pi_{\Gamma \Gamma}(\om)$, included for completeness, describes the electronic dressing of the phonon that is already present in the absence of the cavity. The light--matter interaction underlying phonon-polariton formation is entirely encoded in the electronic current-current response, $\Pi_{jj}(\omega)$, which describes the electromagnetic response of the electronic medium, and the mixed current-phonon response, $\Pi_{j\Gamma}(\omega)$. The latter plays a central role, as it quantifies the electronic current generated by a lattice distortion, thereby determining how efficiently oscillator strength is transferred from the electronic degrees of freedom to the phonon and, ultimately, the strength of the effective phonon--photon coupling. Both response functions contain paramagnetic and diamagnetic contributions, given by~\cite{Note1}
\beq
\Pi_{j \Gamma}(\om) = \chi_{j \Gamma}(\om)  - \chi_{j \Gamma}(0),
\quad
\Pi_{j j}(\om) = \chi_{j j}(\om)  - \chi_{j j}(0).
\label{analytic_cont}
\eeq
Here $\chi_{j \Gamma}(\om)$ is the analytic continuation to real frequencies of $\chi_{j \Gamma}(\mathrm{i}\Om_n) \equiv - \int_0^{\be} d \tau \langle T_{\tau} \hat{j}(\tau) \hat{\Gamma}(0) \rangle e^{\mathrm{i}\Om_n \tau}/N$, where $\tau$ is the imaginary time, $\mathrm{i}\Om_n$ are bosonic Matsubara frequencies, and $\be$ is the inverse temperature. The current--current correlator $\chi_{jj}(\om)$ is defined analogously. An important consequence of Eq.~\eqref{analytic_cont} is that both response functions involving the current operator vanish in the static limit, a property that is crucial for forbidding photon condensation, as discussed below. Physically, this reflects the fact that a static and spatially uniform vector potential cannot induce an electromagnetic response, which is ensured by the gauge-invariant structure of Eq.~\eqref{analytic_cont}. Equation \eqref{2x2} therefore reduces phonon--polariton formation to the competition between electronic screening and dissipation through $\Pi_{jj}(\omega)$ and electron-mediated phonon--photon coupling through $\Pi_{j\Gamma}(\omega)$.

\emph{Electronic response and phonon--photon coupling}---We first consider the current--current response $\Pi_{jj}(\omega)$, which is related to the optical conductivity through
$\Pi_{jj}(\omega)=-\mathrm{i}\omega\sigma(\omega)$ \cite{Mahan2000} and therefore controls both electromagnetic screening and Ohmic dissipation. In a metallic medium, when $\omega_{{\rm c},\mathbf q}$ falls within a frequency range of sizable optical conductivity, the dissipative part of the electronic response broadens the cavity resonance through nonradiative losses, while the reactive part screens and renormalizes the cavity field, detuning the photon mode from the phonon resonance. Dissipation and screening ultimately suppress polariton formation. By contrast, in an insulator, the optical conductivity vanishes below the charge gap $\Delta$, protecting the cavity resonance from electronic damping. This dichotomy between gapless and gapped systems has been demonstrated experimentally~\cite{Jarc2024Multimode}. However, the optical conductivity increases sharply above the gap, and qualitatively different behavior can emerge when the cavity frequency approaches the onset of interband transitions, as discussed below. Electronic excitations thus generally play a detrimental role through $\Pi_{jj}$ by screening and damping the cavity field. Whether polariton formation survives this electronic screening depends on the mixed response $\Pi_{j\Gamma}$ that mediates phonon--photon coupling.

Therefore, a crucial question is whether the low-energy electronic excitations responsible for Drude screening and dissipation also contribute to the electronic response mediating phonon--photon coupling. Remarkably, in the uniform limit they do not. In a time-reversal-invariant system~\cite{morimoto}, we find that the intraband contribution to $\Pi_{j\Gamma}(\omega)$ vanishes even in the presence of impurity scattering~\cite{Note1}. In centrosymmetric systems, the intraband matrix elements vanish identically, whereas in noncentrosymmetric systems they are finite but cancel upon momentum averaging. This cancellation follows from the fact that, in the $q\rightarrow0$ limit, $\hat{\mathbf j}$ is odd under time reversal, while $\hat{\Gamma}_{\al}$ is even. The absence of intraband contributions implies that, unlike $\Pi_{jj}(\om)$, the response $\Pi_{j\Gamma}(\om)$ has no low-energy dissipative Drude contribution, even in a metal. Therefore, its low-energy behavior does not distinguish between gapped and gapless systems: below the onset $\Delta$ of interband transitions, $\Pi_{j\Gamma}(\om)$ is analytic in frequency. Its leading low-frequency behavior, consistent with time reversal symmetry, is thus
\beq
\Pi_{j\Gamma}(\om) \underset{\omega\to 0}{\sim} \mathrm{i}\om Z_0,
\label{pijGsmall}
\eeq
where the prefactor $Z_0$ is directly related to the static Born effective charge of the phonon~\cite{stengel22, sellati2025}, whose value is governed by interband matrix elements. This establishes a fundamental asymmetry between the two electronic responses: intraband excitations can strongly contribute to $\Pi_{jj}$, producing screening and dissipation, while they do not contribute to $\Pi_{j\Gamma}$ in the uniform limit. 

\emph{Generalized no-go theorem}---Before turning to material-dependent consequences, we point out an important implication of Eq.~\eqref{pijGsmall}: cavity photons completely decouple from a critical soft phonon mode, as encountered near, e.g., ferroelectric transitions~\cite{PhysRevX.10.041027}. This extends to tripartite electron--phonon--photon systems existing no-go theorems for photon condensation in purely electronic systems, which rely on the TRK sum rule and the resulting condition $\Pi_{jj}(\omega\rightarrow0)=0$~\cite{Nataf2010,Andolina2019,Andolina2020,Guerci2020,RomanRoche2022,Andolina2022NonPerturbative}. Ignoring $\Pi_{\Gamma\Gamma}$ and $\Pi_{jj}$ for simplicity, as appropriate for off-resonant insulators, and using the asymptotic form in Eq.~\eqref{pijGsmall}, the coupled phonon-photon system supports two polariton branches with frequencies given by 
\begin{equation}
\begin{aligned}
2\omega_{\pm,\bq}^2 ={}&
\omega_{\rm c,\bq}^2+\omega_{\rm p,\bq}^2 +C_{\mathbf{q}} \nonumber\\ 
&\pm \sqrt{ \left( \omega_{\rm c,\bq}^2+\omega_{\rm p,\bq}^2
+C_{\mathbf{q}} \right)^2 -4\omega_{\rm c,\bq}^2\omega_{\rm p,\bq}^2 },
\label{polariton_branches}
\end{aligned}
\end{equation}
where $C_{\mathbf{q}}\equiv4g_{\bq}^2 Z_0^2\lambda^2
\omega_{\rm c,\bq}\omega_{\rm p,\bq}$. The photon propagator, given by the matrix element $\hat M^{-1}_{11}$, 
has poles at $\omega_{+,\bq}$ and $\omega_{-,\bq}$. The corresponding photon spectral weights of the upper and lower polaritons are
\beq
A_{+,\bq}
=
\frac{\omega_{+,\bq}^2-\omega_{\rm p,\bq}^2}
{\omega_{+,\bq}^2-\omega_{-,\bq}^2},
\qquad
A_{-,\bq}
=
\frac{\omega_{\rm p,\bq}^2-\omega_{-,\bq}^2}
{\omega_{+,\bq}^2-\omega_{-,\bq}^2}.
\eeq
We now let the bare phonon soften by taking $\omega_{\rm p,\bq}^2\rightarrow0$. As the lower polariton approaches criticality, its photon weight vanishes as
\begin{equation}
A_{-,\bq}\simeq4 g_{\mathbf{q}}^2 \lambda^2Z_0^2 \left(\frac{\omega_{\rm p,\bq}}{\omega_{\rm c,\bq}}\right)^3,
\end{equation}
such that the critical mode becomes purely phononic. Physically, the cavity field progressively decouples from the soft mode because the effective light--matter coupling $\Pi_{j\Gamma}(\omega)$ vanishes in the static limit, a direct consequence of gauge invariance even in the absence of time-reversal symmetry.

\begin{figure}
\includegraphics[width=0.48\textwidth]{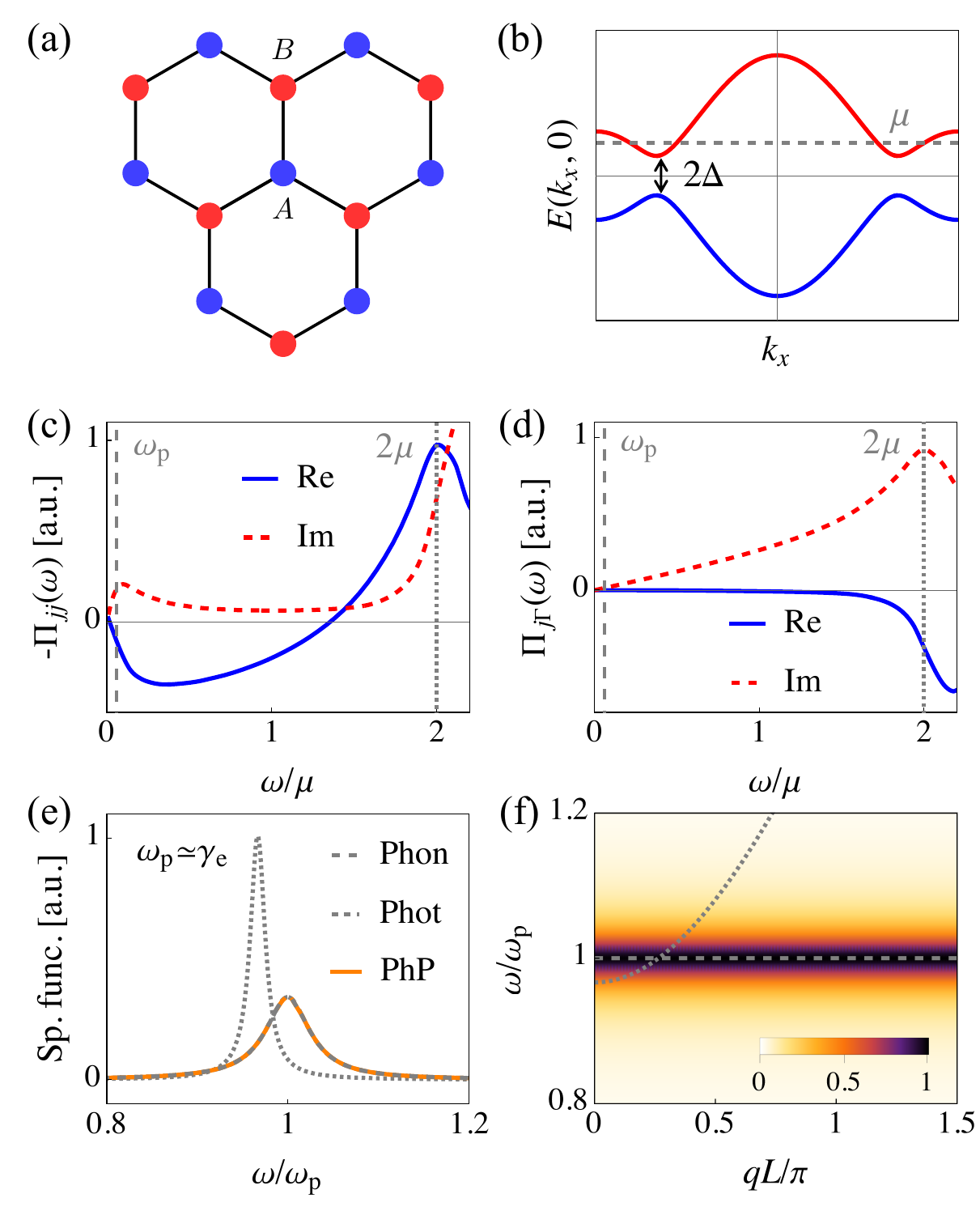}
\caption{(a) Honeycomb lattice with inequivalent sublattices and (b) corresponding band structure along $k_y=0$, showing the chemical potential $\mu$ (dashed line) and band gap $2\Delta$. Real (blue) and imaginary (red dashed) parts of (c) the current-current response $\Pi_{jj}(\omega)$ and (d) the current-phonon response $\Pi_{j\Gamma}(\omega)$ for metallic monolayer NbSe$_2$. Vertical dotted and dashed lines mark the interband threshold $2\mu$ and TO-phonon frequency $\omega_{\rm p}$, respectively. (e) Phonon-polariton spectral function $\mathrm{Im}[\widetilde B_{\mathbf q=0}(\omega+i\delta)]$ (orange), compared with the bare photon $\mathrm{Im}[W_{\mathbf q=0}(\omega+i\delta)]$ (dotted) and cavity-free dressed phonon $\mathrm{Im}[B_{\mathbf q=0}^{-1}(\omega+i\delta)-\lambda^2\Pi_{\Gamma\Gamma}(\omega)]^{-1}$ (dashed), with $\delta=0.01\omega_{\rm p}$. (f) Momentum-resolved phonon-polariton spectral function, normalized to its maximum. Dotted and dashed lines indicate the bare photon dispersion $\omega_{\rm c,\mathbf q}$ and phonon frequency $\omega_{\rm p}$, respectively. Model parameters are given in the SM~\cite{Note1}.} 
\label{Fig1}
\end{figure}

\emph{Microscopic model}---We now turn to the finite-frequency regime and illustrate how the electronic structure controls phonon--polariton formation within a microscopic model. We consider an effective tight-binding Hamiltonian of spinless electrons on a honeycomb lattice with two inequivalent atoms per unit cell [Fig.~\ref{Fig1}(a)]~\cite{Note1},
\[
\ham_{\rm e} =\sum_{\mathbf{k}} \begin{pmatrix} c_{A,\mathbf{k}}^{\dagger} & c_{B,\mathbf{k}}^{\dagger} \end{pmatrix} \begin{pmatrix} 
\Delta-\mu & -t\epsilon_{\mathbf{k}} \\ -t\epsilon_{\mathbf{k}}^* &-\Delta-\mu \end{pmatrix} \begin{pmatrix} c_{A,\mathbf{k}} \\ c_{B,\mathbf{k}} \end{pmatrix}.
\]
Here, $\epsilon_{\mathbf{k}}$ is the dimensionless electronic dispersion, $t$ is the nearest neighbor hopping energy, $\mu$ the chemical potential and $\Delta$ the sublattice energy difference. We focus on the zone-centre doubly degenerate in-plane transverse optical (TO) phonons, assumed to be dispersionless with frequency $\omega_{\mathrm{p}}$. As detailed in the SM~\cite{Note1}, these modes couple to the electrons through the modulation of the nearest-neighbor hopping induced by the lattice displacement $Q_{\alpha}$  along $\alpha=(x,y)$ directions~\cite{guinea07,fachin24}. For $\Delta\ne0$, the inequivalence of the $A$ and $B$ sublattices breaks inversion symmetry and makes the TO phonon infrared active, thereby enabling its coupling to the cavity photon field. In what follows, we focus on the $\alpha=x$ mode and omit Cartesian indices for brevity. We extend the model to an effective quasi-2D material of thickness $d$, consisting of $n_z$ decoupled layers stacked along $z$ and centered within the cavity. This amounts to rescaling the electron--photon coupling as $U_{\bq}\rightarrow U_{\bq}(d/a_z)^{1/2}$, where $a_z$ is the interlayer spacing along $z$~\cite{Note1}. The resulting coupling scales as $U_{\bq}\propto\sqrt{\zeta}$, with $\zeta=d/L$ the cavity filling factor, consistent with experiments~\cite{BarraBurillo2021}.

The balance between screening, dissipation and phonon--photon coupling depends crucially on the position of the cavity and phonon frequencies relative to the electronic excitation spectrum. We therefore consider successively three regimes within the same model: a metal dominated by low-energy intraband screening, a wide-gap insulator in which electronic dissipation is suppressed, and a semiconductor in which the phonon is resonant with interband excitations.

We first consider the regime $\mu>\Delta$ [Fig.~\ref{Fig1}(b)], in which the model provides an effective description of odd-layer metallic NbSe$_2$ thin films, where inversion symmetry is broken~\cite{habara,phonNbSe2}. As detailed in the SM, we evaluate the electronic correlators in the presence of a sizable electronic broadening $\gamma_{\rm e}$ due to impurities~\cite{udina26}. The current-current response $\Pi_{jj}(\omega)$ acquires a finite intraband contribution at $\omega\sim\gamma_{\rm e}$, directly associated with the Drude conductivity~\cite{basov2001}, whereas interband processes contribute only at higher frequencies, $\omega\gtrsim2\mu$ [Fig.~\ref{Fig1}(c)]. The phonon-phonon correlator $\Pi_{\Gamma\Gamma}(\omega)$ shows a similar structure.  
For simplicity, its real part is effectively absorbed into the phonon frequency $\omega_{\rm p}$. 
By contrast, $\Pi_{j\Gamma}(\omega)$ has no intraband contribution and its low-frequency behavior is controlled by off-resonant interband transitions. As a result, the current-phonon correlator is purely imaginary and linear in frequency [Fig.~\ref{Fig1}(d)], in agreement with Eq.~\eqref{pijGsmall}. We next substitute these correlators in Eq.~\eqref{2x2} and obtain the phonon propagator from the matrix element $\hat M^{-1}_{22}$, i.e.\ $\widetilde B_{\mathbf{q}}(\omega)^{-1}\equiv
B_{\mathbf{q}}(\omega)^{-1}-
\left[
\lambda^2\Pi_{\Gamma\Gamma}(\omega)
+
\Sigma^{B}(\omega,\mathbf{q})
\right]$, with the cavity-induced phonon self-energy
\[
\Sigma^B(\omega,\mathbf{q})= \frac{2\lambda^2 g^2_{\mathbf{q}} \omega_{\rm c,\mathbf{q}}\Pi_{j\Gamma}(\omega)\Pi_{ j\Gamma}(-\omega)}{\omega^2-\omega_{\rm c,\mathbf{q}}^2- 2 g^2_{\mathbf{q}} \omega_{\rm c,\mathbf{q}} \Pi_{jj}(\omega)}.
\]
For $\omega_{\rm p}\simeq\omega_{\rm c,0}\simeq\gamma_{\rm e}$, the large dissipative response encoded in $\operatorname{Im}[\Pi_{\Gamma\Gamma}(\omega)]$ and $\operatorname{Im}[\Pi_{jj}(\omega)]$ strongly broadens the bare phonon and cavity modes, respectively, while the sizable reactive contribution $\operatorname{Re}[\Pi_{jj}(\omega)]$ screens and renormalizes the cavity field. Together, dissipation and screening suppress the light--matter hybridization ($\Sigma^B\simeq 0$) and prevent a resolvable splitting of the polariton branches in NbSe$_2$ [Fig.~\ref{Fig1}(e)--(f)], even when the number of layers is increased to enhance the effective light--matter coupling. The metallic regime therefore exposes the detrimental role of low-energy electronic screening and dissipation. 

\begin{figure}
\includegraphics[width=0.48\textwidth]{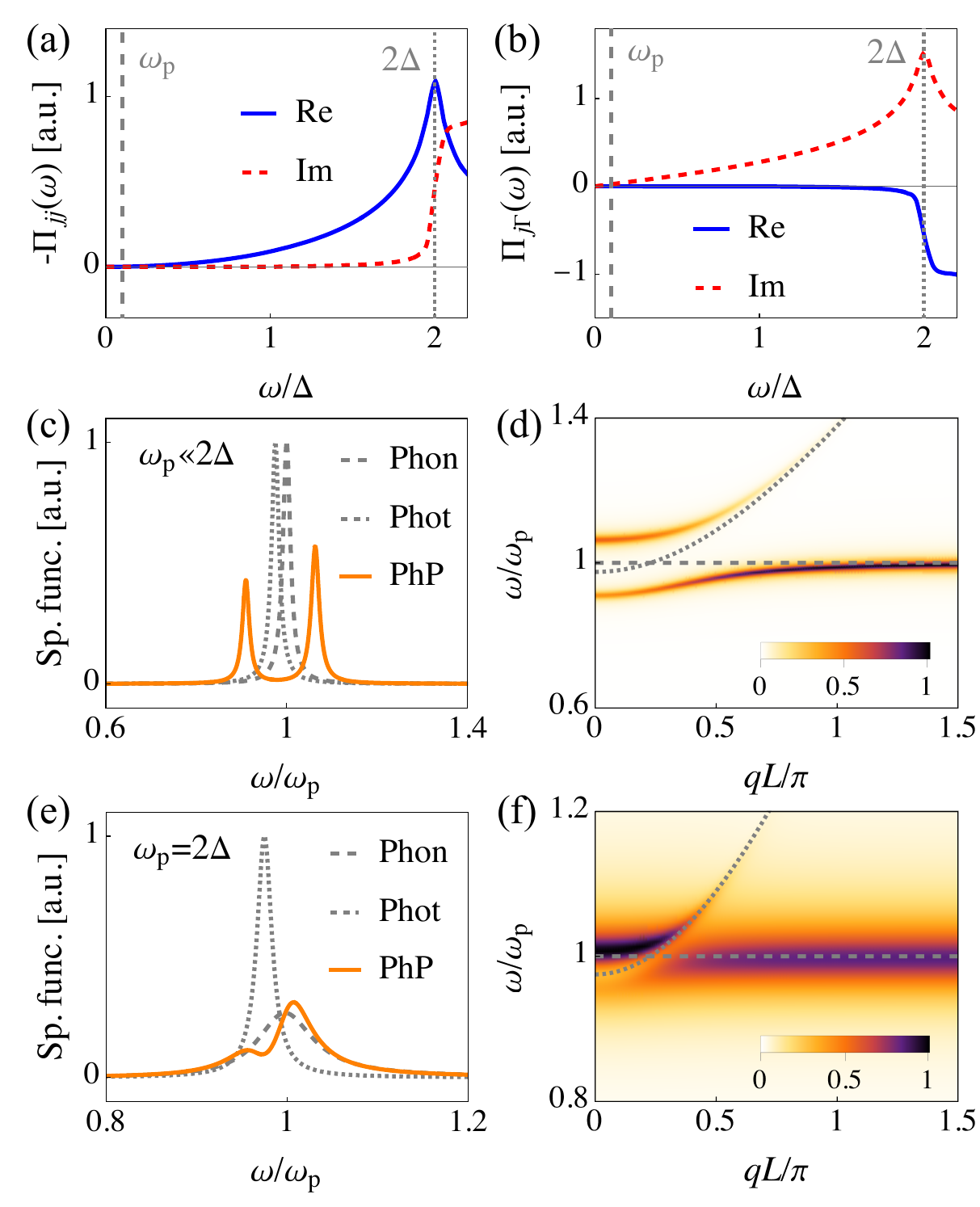}
\caption{Same as Fig.~\ref{Fig1}, but for a film of wide-gap insulating hBN with $\omega_{\rm p}\ll 2\Delta$ [(a)--(d)], and monolayer gapped graphene with $\omega_{\rm p}=2\Delta$ [(e)--(f)]. Vertical dashed and dotted lines mark the phonon frequency $\omega_{\rm p}$ and interband threshold $2\Delta$, respectively. Model parameters are given in the SM~\cite{Note1}.} 
\label{Fig2}
\end{figure}

To switch off these effects, we next remove intraband excitations by setting $\mu=0$ and choosing a sufficiently large $\Delta$, corresponding to wide-gap insulating hBN. In this regime, the intraband contribution to the current--current correlator vanishes as well. Because the cavity photon and TO phonon lie well below the electronic gap, $\omega_{\rm p}\simeq\omega_{\rm c,0}\ll2\Delta$, both $\Pi_{\Gamma\Gamma}(\omega)$ and $\Pi_{jj}(\omega)$ are strongly suppressed at the relevant frequencies [Fig.~\ref{Fig2}(a)], making electronic screening and absorption negligible. By contrast, $\Pi_{j\Gamma}(\omega)$ retains the same low-frequency structure as in the metallic case [Fig.~\ref{Fig2}(b)]. The resulting light-matter hybridization produces a sizable splitting of the phonon-polariton branches [Fig.~\ref{Fig2}(c)-(d)], which increases with the sample thickness, in agreement with experiments~\cite{BarraBurillo2021}. 

The wide-gap limit suppresses electronic losses, but also places electronic excitations far off resonance. We therefore ask whether approaching an interband resonance can enhance the electron-mediated coupling sufficiently to overcome the absorption it introduces. To address this regime, we reduce the gap such that $2\Delta\simeq\omega_{\rm p}$, as relevant to gapped graphene \cite{Bistoni_2019}. Both $\Pi_{jj}(\omega)$ and $\Pi_{\Gamma\Gamma}(\omega)$ then acquire sizable imaginary parts at the phonon frequency, broadening the cavity photon and bare phonon modes, respectively. Crucially, the same interband processes responsible for this absorption strongly enhance $\Pi_{j\Gamma}(\omega)$ near resonance, where it develops sizable real and imaginary parts~\cite{cappelluti, Bistoni_2019}. This has two consequences. First, the enhanced $\mathrm{Im}[\Pi_{j\Gamma} (\omega)]$ can overcome the absorption effect of the cavity field associated with $\mathrm{Im}[\Pi_{jj}(\omega)]$, allowing well-defined polariton branches to emerge [Fig.~2(e)-(f)]. Second, because $\Pi_{j\Gamma}$ acquires a finite real part, the imaginary part of the cavity-induced phonon self-energy $\Sigma^{B}$ contains an additional contribution that partially reduces the phonon damping present in the absence of the cavity field. This effect is reminiscent of Friedrich--Wintgen-like linewidth narrowing~\cite{friedrich}, although here it arises from the complex structure of the electronic response rather than from interference between two independent decay channels. Importantly, this resonant enhancement shifts the balance between hybridization and dissipation: despite a smaller polariton splitting than in multilayer hBN, monolayer gapped graphene exhibits much stronger hybridization than metallic NbSe$_2$ due to its proximity to the interband resonance. This establishes electronic resonance engineering as a route to enhance light--matter hybridization despite screening and absorption.

\emph{Conclusion}---We have developed a gauge-invariant microscopic theory of cavity phonon polaritons, revealing how the interplay of electron-mediated coupling, screening, and dissipation suppresses hybridization in metals while enabling robust polaritons in gapped systems, particularly near interband resonances. Independently of these material-specific effects, gauge invariance enforces photon decoupling from a critical soft phonon, forbidding photon condensation. Our results provide a microscopic framework for understanding cavity-modified lattice dynamics and interpreting spectroscopic experiments, opening new routes toward engineering phonon--polariton formation in quantum materials through electronic resonances.

\emph{Acknowledgment}--- We acknowledge useful discussions with F.\ Macheda, J.\ Fiore, P.\ Fachin, G.\ Marchese, L.\ Giacomelli, Y.\ Laplace, N.\ Sellati, C.\ Ciuti, M.\ Schirò, O.\ Dmytruk. This work was supported by the French National Research Agency (ANR) under Grant No. ANR-23-CE30-0030 (``SUPER2DTMD'').

\footnotetext[1]{See Supplemental Material for the derivation of the phonon-polariton propagator; its extension to a multilayer geometry; the proof of the absence of intraband contributions to the mixed current-phonon response and its relation to time-reversal symmetry; and details of the calculations for the microscopic honeycomb-lattice model.}

\bibliography{references}

@article{sellati2025,
	title = {Theory of terahertz pump optical probe spectroscopy of phonon polaritons in noncentrosymmetric systems},
	volume = {10},
	rights = {2025 The Author(s)},
	issn = {2397-4648},
	url = {https://www.nature.com/articles/s41535-025-00761-8},
	doi = {10.1038/s41535-025-00761-8},
	pages = {46},
	number = {1},
	journal = {npj Quantum Materials},
	publisher = {Nature Publishing Group},
	author = {Sellati, NiccolÃ² and Fiore, Jacopo and Villani, Stefano Paolo and Benfatto, Lara and Udina, Mattia},
	urldate = {2025-12-17},
	year = {2025},
	langid = {english},
}

@article{resta,
  title = {Quantum-Mechanical Position Operator in Extended Systems},
  author = {Resta, Raffaele},
  journal = {Phys. Rev. Lett.},
  volume = {80},
  issue = {9},
  pages = {1800--1803},
  numpages = {0},
  year = {1998},
  month = {Mar},
  publisher = {American Physical Society},
  doi = {10.1103/PhysRevLett.80.1800},
  url = {https://link.aps.org/doi/10.1103/PhysRevLett.80.1800}
}

@article{habara,
  title = {Symmetry manipulation of nonlinear optical effect for metallic transition-metal dichalcogenides},
  author = {Habara, Ren and Wakabayashi, Katsunori},
  journal = {Phys. Rev. B},
  volume = {107},
  issue = {11},
  pages = {115422},
  numpages = {18},
  year = {2023},
  month = {Mar},
  publisher = {American Physical Society},
  doi = {10.1103/PhysRevB.107.115422},
  url = {https://link.aps.org/doi/10.1103/PhysRevB.107.115422}
}

@article{morimoto,
	title = {Nonreciprocal current from electron interactions in noncentrosymmetric crystals: roles of time reversal symmetry and dissipation},
	volume = {8},
	issn = {2045-2322},
	url = {https://doi.org/10.1038/s41598-018-20539-2},
	doi = {10.1038/s41598-018-20539-2},
	pages = {2973},
	number = {1},
	journal = {Scientific Reports},
	author = {Morimoto, Takahiro and Nagaosa, Naoto},
	year = 2018,
}

@article{Nguyen2023,
  author  = {Nguyen, Danh-Phuong and Arwas, Geva
             and Lin, Zuzhang and Yao, Wang
             and Ciuti, Cristiano},
  title   = {Electron-Photon Chern Number in
             Cavity-Embedded 2D Moir{\'e} Materials},
  journal = {Physical Review Letters},
  volume  = {131},
  number  = {17},
  pages   = {176602},
  year    = {2023},
  doi     = {10.1103/PhysRevLett.131.176602},
  url     = {https://doi.org/10.1103/PhysRevLett.131.176602}
}

@article{Nataf2010,
  author  = {Nataf, Pierre and Ciuti, Cristiano},
  title   = {No-go theorem for superradiant quantum phase transitions in cavity QED and counter-example in circuit QED},
  journal = {Nature Communications},
  volume  = {1},
  pages   = {72},
  year    = {2010},
  doi     = {10.1038/ncomms1069}
}

@article{PhysRevA.98.053819,
  title = {Breakdown of gauge invariance in ultrastrong-coupling cavity QED},
  author = {De Bernardis, Daniele and Pilar, Philipp and Jaako, Tuomas and De Liberato, Simone and Rabl, Peter},
  journal = {Phys. Rev. A},
  volume = {98},
  issue = {5},
  pages = {053819},
  numpages = {15},
  year = {2018},
  month = {Nov},
  publisher = {American Physical Society},
  doi = {10.1103/PhysRevA.98.053819},
  url = {https://link.aps.org/doi/10.1103/PhysRevA.98.053819}
}

@article{Brar2014,
  author  = {Brar, Victor W. and Jang, Min Seok and Sherrott, Michelle and Kim, Seyoon and Lopez, Javier J. and Kim, Laura B. and Choi, Min and Atwater, Harry A.},
  title   = {Hybrid Surface-Phonon-Plasmon Polariton Modes in Graphene/Monolayer h-BN Heterostructures},
  journal = {Nano Letters},
  volume  = {14},
  number  = {7},
  pages   = {3876--3880},
  year    = {2014},
  doi     = {10.1021/nl501096s},
  url     = {https://doi.org/10.1021/nl501096s}
}

@article{phonNbSe2,
  title = {Comprehensive optical characterization of atomically thin ${\mathrm{NbSe}}_{2}$},
  author = {Hill, Heather M. and Rigosi, Albert F. and Krylyuk, Sergiy and Tian, Jifa and Nguyen, Nhan V. and Davydov, Albert V. and Newell, David B. and Hight Walker, Angela R.},
  journal = {Phys. Rev. B},
  volume = {98},
  issue = {16},
  pages = {165109},
  numpages = {7},
  year = {2018},
  month = {Oct},
  publisher = {American Physical Society},
  doi = {10.1103/PhysRevB.98.165109},
  url = {https://link.aps.org/doi/10.1103/PhysRevB.98.165109}
}

@article{basov2001,
  title = {Anisotropic electrodynamics of layered metal $2H\ensuremath{-}{\mathrm{NbSe}}_{2}$},
  author = {Dordevic, S. V. and Basov, D. N. and Dynes, R. C. and Bucher, E.},
  journal = {Phys. Rev. B},
  volume = {64},
  issue = {16},
  pages = {161103(R)},
  numpages = {4},
  year = {2001},
  month = {Sep},
  publisher = {American Physical Society},
  doi = {10.1103/PhysRevB.64.161103},
  url = {https://link.aps.org/doi/10.1103/PhysRevB.64.161103}
}

@article{udina26,
  title = {Antisymmetric Raman Response},
  author = {Udina, Mattia and Paul, Indranil},
  journal = {Phys. Rev. Lett.},
  volume = {136},
  issue = {12},
  pages = {126505},
  numpages = {6},
  year = {2026},
  month = {Mar},
  publisher = {American Physical Society},
  doi = {10.1103/gnpt-5ffq},
  url = {https://link.aps.org/doi/10.1103/gnpt-5ffq}
}

@article{fachin24,
  title = {Nearly quantized Born effective charges as probes for the topological phase transition in the Haldane and Kane-Mele models},
  author = {Fachin, Paolo and Macheda, Francesco and Barone, Paolo and Mauri, Francesco},
  journal = {Phys. Rev. B},
  volume = {110},
  issue = {20},
  pages = {L201405},
  numpages = {7},
  year = {2024},
  month = {Nov},
  publisher = {American Physical Society},
  doi = {10.1103/PhysRevB.110.L201405},
  url = {https://link.aps.org/doi/10.1103/PhysRevB.110.L201405}
}

@article{friedrich,
  title = {Interfering resonances and bound states in the continuum},
  author = {Friedrich, H. and Wintgen, D.},
  journal = {Phys. Rev. A},
  volume = {32},
  issue = {6},
  pages = {3231--3242},
  numpages = {0},
  year = {1985},
  month = {Dec},
  publisher = {American Physical Society},
  doi = {10.1103/PhysRevA.32.3231},
  url = {https://link.aps.org/doi/10.1103/PhysRevA.32.3231}
}

@article{Bistoni_2019,
doi = {10.1088/2053-1583/ab2ce0},
url = {https://doi.org/10.1088/2053-1583/ab2ce0},
year = {2019},
month = {jul},
publisher = {IOP Publishing},
volume = {6},
number = {4},
pages = {045015},
author = {Bistoni, O and Barone, P and Cappelluti, E and Benfatto, L and Mauri, F},
title = {Giant effective charges and piezoelectricity in gapped graphene},
journal = {2D Materials}
}

@article{cappelluti,
  title = {Charged-phonon theory and Fano effect in the optical spectroscopy of bilayer graphene},
  author = {Cappelluti, E. and Benfatto, L. and Manzardo, M. and Kuzmenko, A. B.},
  journal = {Phys. Rev. B},
  volume = {86},
  issue = {11},
  pages = {115439},
  numpages = {15},
  year = {2012},
  month = {Sep},
  publisher = {American Physical Society},
  doi = {10.1103/PhysRevB.86.115439},
  url = {https://link.aps.org/doi/10.1103/PhysRevB.86.115439}
}

@article{binci21,
  title = {First-principles theory of infrared vibrational spectroscopy of metals and semimetals: Application to graphite},
  author = {Binci, Luca and Barone, Paolo and Mauri, Francesco},
  journal = {Phys. Rev. B},
  volume = {103},
  issue = {13},
  pages = {134304},
  numpages = {10},
  year = {2021},
  month = {Apr},
  publisher = {American Physical Society},
  doi = {10.1103/PhysRevB.103.134304},
  url = {https://link.aps.org/doi/10.1103/PhysRevB.103.134304}
}

@article{guinea07,
  title = {Electron-phonon coupling and Raman spectroscopy in graphene},
  author = {Castro Neto, A. H. and Guinea, Francisco},
  journal = {Phys. Rev. B},
  volume = {75},
  issue = {4},
  pages = {045404},
  numpages = {8},
  year = {2007},
  month = {Jan},
  publisher = {American Physical Society},
  doi = {10.1103/PhysRevB.75.045404},
  url = {https://link.aps.org/doi/10.1103/PhysRevB.75.045404}
}

@article{stengel22,
  title = {Nonadiabatic Born Effective Charges in Metals and the Drude Weight},
  author = {Dreyer, Cyrus E. and Coh, Sinisa and Stengel, Massimiliano},
  journal = {Phys. Rev. Lett.},
  volume = {128},
  issue = {9},
  pages = {095901},
  numpages = {6},
  year = {2022},
  month = {Feb},
  publisher = {American Physical Society},
  doi = {10.1103/PhysRevLett.128.095901},
  url = {https://link.aps.org/doi/10.1103/PhysRevLett.128.095901}
}

@article{Jarc2024Multimode,
  author  = {Giacomo Jarc and Shahla Yasmin Mathengattil and
             Angela Montanaro and Enrico Maria Rigoni and
             Simone Dal Zilio and Daniele Fausti},
  title   = {Multimode vibrational coupling across the insulator-to-metal transition in 1T-TaS$_2$ in THz cavities},
  journal = {The Journal of Chemical Physics},
  volume  = {161},
  number  = {15},
  pages   = {154711},
  year    = {2024},
  doi     = {10.1063/5.0231533},
  url     = {https://doi.org/10.1063/5.0231533}
}

@article{Kozin2025,
  title = {Cavity-enhanced superconductivity via band engineering},
  author = {Kozin, Valerii K. and Thingstad, Even and Loss, Daniel and Klinovaja, Jelena},
  journal = {Phys. Rev. B},
  volume = {111},
  issue = {3},
  pages = {035410},
  numpages = {9},
  year = {2025},
  month = {Jan},
  publisher = {American Physical Society},
  doi = {10.1103/PhysRevB.111.035410},
  url = {https://link.aps.org/doi/10.1103/PhysRevB.111.035410}
}

@article{Lu2024CavityEnhanced,
  author  = {Lu, I-Te and Shin, Dongbin and Svendsen, Mark Kamper and H{\"u}bener, Hannes and De Giovannini, Umberto and Latini, Simone and Ruggenthaler, Michael and Rubio, Angel},
  title   = {Cavity-enhanced superconductivity in {MgB$_2$} from first-principles quantum electrodynamics ({QEDFT})},
  journal = {Proc. Natl. Acad. Sci. U.S.A.},
  volume  = {121},
  number  = {50},
  pages   = {e2415061121},
  year    = {2024},
  doi     = {10.1073/pnas.2415061121},
  url     = {https://doi.org/10.1073/pnas.2415061121}
}

@article{Andolina2020,
  author       = {Andolina, G. M. and Pellegrino, F. M. D. and Giovannetti, V. and MacDonald, A. H. and Polini, M.},
  title        = {Theory of photon condensation in a spatially varying electromagnetic field},
  journal      = {Phys. Rev. B},
  volume       = {102},
  number       = {12},
  pages        = {125137},
  year         = {2020},
  publisher    = {American Physical Society},
  doi          = {10.1103/PhysRevB.102.125137}
}

@article{Jiang2026NoGo,
  author        = {Jiang, Qing-Dong},
  title         = {No-Go Theorem and Routes towards Cavity-Enhanced Superconductivity},
  journal       = {arXiv preprint arXiv:2608.14784},
  year          = {2026},
  eprint        = {2608.14784},
  archivePrefix = {arXiv},
  primaryClass  = {cond-mat.supr-con},
  doi           = {10.48550/arXiv.2608.14784},
  url           = {https://arxiv.org/abs/2608.14784}
}

@article{Riolo2025Tuning,
  author  = {Riolo, Riccardo and Tomadin, Andrea and Mazza, Giacomo and
             Asgari, Reza and MacDonald, Allan H. and Polini, Marco},
  title   = {Tuning Fermi liquids with polaritonic cavities},
  journal = {Proc. Natl. Acad. Sci. U.S.A.},
  volume  = {122},
  number  = {32},
  pages   = {e2407995122},
  year    = {2025},
  doi     = {10.1073/pnas.2407995122},
  url     = {https://doi.org/10.1073/pnas.2407995122}
}

@article{AndolinaAmperean2024,
  title = {Amperean superconductivity cannot be induced by deep subwavelength cavities in a two-dimensional material},
  author = {Andolina, Gian Marcello and De Pasquale, Antonella and Pellegrino, Francesco Maria Dimitri and Torre, Iacopo and Koppens, Frank H. L. and Polini, Marco},
  journal = {Phys. Rev. B},
  volume = {109},
  issue = {10},
  pages = {104513},
  numpages = {22},
  year = {2024},
  month = {Mar},
  publisher = {American Physical Society},
  doi = {10.1103/PhysRevB.109.104513},
  url = {https://link.aps.org/doi/10.1103/PhysRevB.109.104513}
}

@article{Helmrich2026,
  author  = {Helmrich, F. and Adlong, H. S. and Kroner, M. and Khanonkin, I. and Scalari, G. and Faist, J. and Imamo{\u{g}}lu, A. and Nova, T. F.},
  title   = {Cavity-driven attractive interactions in quantum materials},
  journal = {Nature},
  volume  = {654},
  pages   = {361--368},
  year    = {2026},
  doi     = {10.1038/s41586-026-10609-1}
}

@article{Curtis2019,
  title = {Cavity Quantum Eliashberg Enhancement of Superconductivity},
  author = {Curtis, Jonathan B. and Raines, Zachary M. and Allocca, Andrew A. and Hafezi, Mohammad and Galitski, Victor M.},
  journal = {Phys. Rev. Lett.},
  volume = {122},
  issue = {16},
  pages = {167002},
  numpages = {5},
  year = {2019},
  month = {Apr},
  publisher = {American Physical Society},
  doi = {10.1103/PhysRevLett.122.167002},
  url = {https://link.aps.org/doi/10.1103/PhysRevLett.122.167002}
}

@article{Schlawin2019,
  title = {Cavity-Mediated Electron-Photon Superconductivity},
  author = {Schlawin, Frank and Cavalleri, Andrea and Jaksch, Dieter},
  journal = {Phys. Rev. Lett.},
  volume = {122},
  issue = {13},
  pages = {133602},
  numpages = {6},
  year = {2019},
  month = {Apr},
  publisher = {American Physical Society},
  doi = {10.1103/PhysRevLett.122.133602},
  url = {https://link.aps.org/doi/10.1103/PhysRevLett.122.133602}
}

@article{Wang2026VacuumEnhanced,
  author  = {Wang, Zheyan and Cardoso, Gabriel and Yang, Liu and Gong, Xun
             and Zhang, Chi and Zhu, Yufei and Zhang, Dongbo and Pan, Nan
             and Cai, Hongbing and Chen, Yong P. and Jiang, Qing-Dong
             and Cheng, Guanghui and Wilczek, Frank and Zeng, Changgan},
  title   = {Evidence for vacuum-enhanced superconductivity in {NbSe$_2$}},
  journal = {Nature},
  year    = {2026},
  doi     = {10.1038/s41586-026-11037-x},
  url     = {https://doi.org/10.1038/s41586-026-11037-x}
}

@article{Thomas2025Exploring,
  author  = {Thomas, A. and Devaux, E. and Nagarajan, K. and
             Chervy, T. and Seidel, M. and Rogez, G. and
             Robert, J. and Drillon, M. and Ruan, T. T. and
             Schlittenhardt, S. and Ruben, M. and
             Hagenm{\"u}ller, D. and Sch{\"u}tz, S. and
             Schachenmayer, J. and Genet, C. and
             Pupillo, G. and Ebbesen, T. W.},
  title   = {Exploring superconductivity under strong coupling with the vacuum electromagnetic field},
  journal = {The Journal of Chemical Physics},
  volume  = {162},
  number  = {13},
  pages   = {134701},
  year    = {2025},
  url = {https://doi.org/10.1063/5.0231202},
  doi     = {10.1063/5.0231202}
}

@article{Sentef2018CavityQED,
  author  = {Sentef, Michael A. and Ruggenthaler, Michael and Rubio, Angel},
  title   = {Cavity quantum-electrodynamical polaritonically enhanced electron-phonon coupling and its influence on superconductivity},
  journal = {Science Advances},
  volume  = {4},
  number  = {11},
  pages   = {eaau6969},
  year    = {2018},
  url = {https://doi.org/10.1126/sciadv.aau6969},
  doi     = {10.1126/sciadv.aau6969}
}

@article{Zhang2026CavityEnhanced,
  author        = {Zhang, Hanxiang and Feng, Zexin and Lu, I-Te and
                   Li, Zhiwei and Guo, Songhao and Shang, Qiuyu and
                   Luan, Dening and Panmai, Mingcheng and
                   Watanabe, Kenji and Taniguchi, Takashi and
                   Rubio, Angel and Gao, Weibo},
  title         = {Cavity Enhanced Superconductivity},
  year          = {2026},
  journal = {arXiv preprint},
  eprint = {arXiv:2606.19171},   
  url           = {https://arxiv.org/abs/2606.19171}
}

@article{Keren2026CavityAltered,
  author  = {Keren, Itai and Webb, Tatiana A. and Zhang, Shuai and
             Xu, Jikai and Sun, Dihao and Kim, Brian S. Y. and
             Shin, Dongbin and Zhang, Songtian S. and Zhang, Junhe and
             Pereira, Giancarlo and Yao, Juntao and Okugawa, Takuya and
             Michael, Marios H. and Vi{\~n}as Bostr{\"o}m, Emil and
             Edgar, James H. and Wolf, Stuart and Julian, Matthew and
             Prasankumar, Rohit P. and Miyagawa, Kazuya and
             Kanoda, Kazushi and Gu, Genda and Cothrine, Matthew and
             Mandrus, David and Buzzi, Michele and Cavalleri, Andrea and
             Dean, Cory R. and Kennes, Dante M. and Millis, Andrew J. and
             Li, Qiang and Sentef, Michael A. and Rubio, Angel and
             Pasupathy, Abhay N. and Basov, D. N.},
  title   = {Cavity-altered superconductivity},
  journal = {Nature},
  volume  = {650},
  number  = {8103},
  pages   = {864--868},
  year    = {2026},
  doi     = {10.1038/s41586-025-10062-6},
  url     = {https://doi.org/10.1038/s41586-025-10062-6}
}

@article{Xu2026VacuumDressed,
  author        = {Xu, Hongjing and Baydin, Andrey and Yi, Qinyan and
                   Lu, I-Te and Zhu, Ningxu and Kritzell, T. Elijah and
                   Doumani, Jacques and Kim, Dasom and Tay, Fuyang and
                   Rubio, Angel and Kono, Junichiro},
  title         = {Vacuum-dressed superconductivity in NbN observed in a high-$Q$ terahertz cavity},
  year          = {2026},
  journal = {arXiv preprint},
  eprint = {arXiv:2601.08191},    
  url           = {https://arxiv.org/abs/2601.08191}
}

@article{Cottet2015ElectronPhoton,
  author  = {Cottet, Audrey and Kontos, Takis and Dou\c{c}ot, Beno\^{\i}t},
  title   = {Electron-photon coupling in mesoscopic quantum electrodynamics},
  journal = {Physical Review B},
  volume  = {91},
  number  = {20},
  pages   = {205417},
  year    = {2015},
  url = {https://doi.org/10.1103/PhysRevB.91.205417},
  doi     = {10.1103/PhysRevB.91.205417}
}

@article{Montanaro2026CavityEnhanced,
  author        = {Montanaro, Angela and Plastovets, Vadim and
                   Khatiwada, Nitesh and Fiore, Jacopo and
                   Jarc, Giacomo and Alabbadi, Abdullah and
                   Mastropasqua, Antonio and Rigoni, Enrico Maria and
                   Mathengattil, Shahla Y. and Dal Zilio, Simone and
                   Fassioli Olsen, Francesca and Novelli, Fabio and
                   Winnerl, Stephan and Sentef, Michael A. and
                   Kennes, Dante M. and Millis, Andrew J. and
                   Piazza, Francesco and Fausti, Daniele},
  title         = {Cavity-enhanced superconducting response in an underdoped cuprate},
  year          = {2026},
  journal = {arXiv preprint},
  eprint = {arXiv:2606.18084},  
  url = {https://arxiv.org/abs/2606.18084},
}

@article{Bretscher2026FluctuationEngineering,
  title         = {Fluctuation engineering in cavity quantum materials},
  author        = {Bretscher, Hope M. and Graziotto, Lorenzo and Michael, Marios H.
                   and Montanaro, Angela and Lu, I-Te and Grankin, Andrey
                   and McIver, James W. and Faist, J{\'e}r{\^o}me
                   and Fausti, Daniele and Eckstein, Martin
                   and Ruggenthaler, Michael and Rubio, Angel
                   and Basov, D. N. and Hafezi, Mohammad
                   and Claassen, Martin and Kennes, Dante M.
                   and Sentef, Michael A.},
  year          = {2026},                 
  journal       = {arXiv preprint}, 
  eprint = {arXiv:2604.08666},  
  url = {https://arxiv.org/abs/2604.08666},
}

@article{BarraBurillo2021,
  author  = {Mar{\'i}a Barra-Burillo and Unai Muniain and Sara Catalano and Marta Autore and F{\`e}lix Casanova and Luis E. Hueso and Javier Aizpurua and Ruben Esteban and Rainer Hillenbrand},
  title   = {Microcavity phonon polaritons from the weak to the ultrastrong phonon--photon coupling regime},
  journal = {Nature Communications},
  volume  = {12},
  pages    = {6206},
  year     = {2021},
  doi      = {10.1038/s41467-021-26060-x},
  url      = {https://doi.org/10.1038/s41467-021-26060-x},
  issn     = {2041-1723},
  publisher= {Nature Portfolio}
}

@book{Mahan2000,
  author    = {Mahan, Gerald D.},
  title     = {Many-Particle Physics},
  edition   = {3},
  publisher = {Springer},
  address   = {New York},
  year      = {2000},
  doi       = {10.1007/978-1-4757-5714-9},
  isbn      = {978-0-306-46338-9},
  url       = {https://doi.org/10.1007/978-1-4757-5714-9}
}

@article{Li2020,
  title = {Electromagnetic coupling in tight-binding models for strongly correlated light and matter},
  author = {Li, Jiajun and Golez, Denis and Mazza, Giacomo and Millis, Andrew J. and Georges, Antoine and Eckstein, Martin},
  journal = {Phys. Rev. B},
  volume = {101},
  issue = {20},
  pages = {205140},
  numpages = {16},
  year = {2020},
  month = {May},
  publisher = {American Physical Society},
  doi = {10.1103/PhysRevB.101.205140},
  url = {https://link.aps.org/doi/10.1103/PhysRevB.101.205140}
}

@Article{RomanRoche2022,
	title={{Effective theory for matter in non-perturbative cavity {QED}}},
	author={Juan Rom\'an-Roche and David Zueco},
	journal={SciPost Phys. Lect. Notes},
	pages={50},
	year={2022},
	publisher={SciPost},
	doi={10.21468/SciPostPhysLectNotes.50},
	url={https://scipost.org/10.21468/SciPostPhysLectNotes.50}
}

@article{PhysRevX.10.041027,
  title = {Quantum Electrodynamic Control of Matter: Cavity-Enhanced Ferroelectric Phase Transition},
  author = {Ashida, Yuto and \ifmmode \dot{I}\else \.{I}\fi{}mamo\ifmmode \breve{g}\else \u{g}\fi{}lu, Ata\ifmmode \mbox{\c{c}}\else \c{c}\fi{} and Faist, J\'er\^ome and Jaksch, Dieter and Cavalleri, Andrea and Demler, Eugene},
  journal = {Phys. Rev. X},
  volume = {10},
  issue = {4},
  pages = {041027},
  numpages = {33},
  year = {2020},
  month = {Nov},
  publisher = {American Physical Society},
  doi = {10.1103/PhysRevX.10.041027},
  url = {https://link.aps.org/doi/10.1103/PhysRevX.10.041027}
}

@article{Jochl2026,
  author        = {J{\"o}chl, Elsa and Helmrich, Felix and Lindel, Frieder and Hale, Lucy and Graziotto, Lorenzo and Jarrahi, Mona and Nova, Tobia F. and Faist, J{\'e}r{\^o}me and Scalari, Giacomo},
  title         = {Absence of a Superradiant Phase Transition in Dirac Landau Polaritons},
  journal       = {arXiv preprint},
  year          = {2026},
  eprint        = {arXiv:2605.27319},
  url           = {https://arxiv.org/abs/2605.27319}
}

@article{Zheng2017,
author = {Zheng, D. and Zhang, S. and Deng, Q. and Kang, M. and Nordlander, P. and Xu, H.},
title = {Manipulating Coherent Plasmon-Exciton Interaction in a Single Silver Nanorod on Monolayer {WS}e$_{2}$},
journal = {Nano Lett.},
volume = {17},
number = {6},
pages = {3809-3814},
year = {2017},
doi = {10.1021/acs.nanolett.7b01176},
URL = {https://doi.org/10.1021/acs.nanolett.7b01176},
}

@article{Liu2016,
author = {Liu, W. and Lee, B. and Naylor, C.~H. and Ee, H.-S. and Park, J. and Johnson, A.~T.~C. and Agarwal, R.},
title = {Strong Exciton-Plasmon Coupling in {M}o{S}$_{2}$ Coupled with Plasmonic Lattice},
journal = {Nano Lett.},
volume = {16},
number = {2},
pages = {1262-1269},
year = {2016},
doi = {10.1021/acs.nanolett.5b04588},
URL = {https://doi.org/10.1021/acs.nanolett.5b04588},
}

@article{Stuhrenberg2018,
author = {St\"{u}hrenberg, M. and Munkhbat, B. and Baranov, D.~G. and Cuadra, J. and Yankovich, A.~B. and Antosiewicz, T.~J. and Olsson, E. and Shegai, T.},
title = {Strong Light-Matter Coupling between Plasmons in Individual Gold Bi-pyramids and Excitons in Mono- and Multilayer {WS}e$_{2}$},
journal = {Nano Lett.},
volume = {18},
number = {9},
pages = {5938-5945},
year = {2018},
doi = {10.1021/acs.nanolett.8b02652},
URL = {https://doi.org/10.1021/acs.nanolett.8b02652},
}

@article{Wang2024,
  author  = {Wang, Hongwei and Kumar, Anshuman and Dai, Siyuan and Lin, Xiao and Jacob, Zubin and Oh, Sang-Hyun and Menon, Vinod and Narimanov, Evgenii and Kim, Young Duck and Wang, Jian-Ping and Avouris, Phaedon and Martin-Moreno, Luis and Caldwell, Joshua D. and Low, Tony},
  title   = {Planar hyperbolic polaritons in 2D van der Waals materials},
  journal = {Nature Communications},
  volume  = {15},
  pages   = {69},
  year    = {2024},
  doi     = {10.1038/s41467-023-43992-8},
  url     = {https://doi.org/10.1038/s41467-023-43992-8}
}

@article{Koppens2011GraphenePlasmonics,
  title   = {Graphene Plasmonics: A Platform for Strong Light--Matter Interactions},
  author  = {Koppens, Frank H. L. and Chang, Darrick E. and Garcia de Abajo, F. Javier},
  journal = {Nano Lett.},
  volume  = {11},
  number  = {8},
  pages   = {3370--3377},
  year    = {2011},
  doi     = {10.1021/nl201771h}
}

@article{Liu2015,
  author  = {Liu, Xiaoze and Galfsky, Tal and Sun, Zheng and Xia, Fengnian and Lin, Erh-Chen and Lee, Yi-Hsien and K{\'e}na-Cohen, St{\'e}phane and Menon, Vinod M.},
  title   = {Strong light--matter coupling in two-dimensional atomic crystals},
  journal = {Nature Photonics},
  volume  = {9},
  pages   = {30--34},
  year    = {2015},
  doi     = {10.1038/nphoton.2014.304},
  url     = {https://doi.org/10.1038/nphoton.2014.304}
}

@article{Manzeli2017,
  author  = {Manzeli, Sajedeh and Ovchinnikov, Dmitry and Pasquier, Diego
             and Yazyev, Oleg V. and Kis, Andras},
  title   = {2D transition metal dichalcogenides},
  journal = {Nature Reviews Materials},
  volume  = {2},
  pages   = {17033},
  year    = {2017},
  doi     = {10.1038/natrevmats.2017.33},
  url     = {https://doi.org/10.1038/natrevmats.2017.33}
}

@article{Wilson2021,
  author  = {Wilson, Nathan P. and Yao, Wang and Shan, Jie and Xu, Xiaodong},
  title   = {Excitons and emergent quantum phenomena in stacked 2D semiconductors},
  journal = {Nature},
  volume  = {599},
  pages   = {383--392},
  year    = {2021},
  doi     = {10.1038/s41586-021-03979-1},
  url     = {https://doi.org/10.1038/s41586-021-03979-1}
}

@article{Autore2018,
  author  = {Autore, Marta and Li, Peining and Dolado, Irene and Alfaro-Mozaz, Francisco J. and Esteban, Ruben and Atxabal, Ainhoa and Casanova, F{\`e}lix and Hueso, Luis E. and Alonso-Gonz{\'a}lez, Pablo and Aizpurua, Javier and Nikitin, Alexey Y. and V{\'e}lez, Sa{\"u}l and Hillenbrand, Rainer},
  title   = {Boron nitride nanoresonators for phonon-enhanced molecular vibrational spectroscopy at the strong coupling limit},
  journal = {Light: Science \& Applications},
  volume  = {7},
  pages   = {17172},
  year    = {2018},
  doi     = {10.1038/lsa.2017.172},
  url     = {https://doi.org/10.1038/lsa.2017.172}
}

@article{Li2015,
  author  = {Li, Peining and Lewin, Martin and Kretinin, Andrey V. and Caldwell, Joshua D. and Novoselov, Kostya S. and Taniguchi, Takashi and Watanabe, Kenji and Gaussmann, Fabian and Taubner, Thomas},
  title   = {Hyperbolic phonon-polaritons in boron nitride for near-field optical imaging and focusing},
  journal = {Nature Communications},
  volume  = {6},
  pages   = {7507},
  year    = {2015},
  doi     = {10.1038/ncomms8507},
  url     = {https://doi.org/10.1038/ncomms8507}
}

@article{Caldwell2014,
  author  = {Caldwell, Joshua D. and Kretinin, Andrey V. and Chen, Yiguo and Giannini, Vincenzo and Fogler, Michael M. and Francescato, Yan and Ellis, Chase T. and Tischler, Joseph G. and Woods, Colin R. and Giles, Alexander J. and Hong, Minghui and Watanabe, Kenji and Taniguchi, Takashi and Maier, Stefan A. and Novoselov, Kostya S.},
  title   = {Sub-diffractional volume-confined polaritons in the natural hyperbolic material hexagonal boron nitride},
  journal = {Nature Communications},
  volume  = {5},
  pages   = {5221},
  year    = {2014},
  doi     = {10.1038/ncomms6221},
  url     = {https://doi.org/10.1038/ncomms6221}
}

@article{Dai2014,
  author  = {Dai, S. and Fei, Z. and Ma, Q. and Rodin, A. S. and Wagner, M. and McLeod, A. S. and Liu, M. K. and Gannett, W. and Regan, W. and Watanabe, K. and Taniguchi, T. and Thiemens, M. and Dominguez, G. and Castro Neto, A. H. and Zettl, A. and Keilmann, F. and Jarillo-Herrero, P. and Fogler, M. M. and Basov, D. N.},
  title   = {Tunable phonon polaritons in atomically thin van der Waals crystals of boron nitride},
  journal = {Science},
  volume  = {343},
  number  = {6175},
  pages   = {1125--1129},
  year    = {2014},
  doi     = {10.1126/science.1246833},
  url     = {https://doi.org/10.1126/science.1246833}
}

@article{Nuckolls2024,
  author  = {Nuckolls, Kevin P. and Yazdani, Ali},
  title   = {A microscopic perspective on moir{\'e} materials},
  journal = {Nature Reviews Materials},
  volume  = {9},
  pages   = {460--480},
  year    = {2024},
  doi     = {10.1038/s41578-024-00682-1},
  url     = {https://doi.org/10.1038/s41578-024-00682-1}
}

@article{Liu2025CavityMediated,
  title   = {Cavity-Mediated Electron-Electron Interactions: Renormalizing Dirac States in Graphene},
  author  = {Liu, Hang and Troisi, Francesco and H{\"u}bener, Hannes and Latini, Simone and Rubio, Angel},
  journal = {Science Advances},
  volume  = {11},
  number  = {43},
  pages   = {eadz1855},
  year    = {2025},
  doi     = {10.1126/sciadv.adz1855},
  url     = {https://doi.org/10.1126/sciadv.adz1855}
}

@article{Chen2017,
  author  = {Chen, Yen-Jung and Cain, Jeffrey D. and Stanev, Teodor K. and Dravid, Vinayak P. and Stern, Nathaniel P.},
  title   = {Valley-polarized exciton--polaritons in a monolayer semiconductor},
  journal = {Nature Photonics},
  volume  = {11},
  pages   = {431--435},
  year    = {2017},
  doi     = {10.1038/nphoton.2017.86},
  url     = {https://doi.org/10.1038/nphoton.2017.86}
}

@article{Flatten2016,
  author  = {Flatten, Lucas C. and He, Zhengyu and Coles, David M. and Trichet, Aur{\'e}lien A. P. and Powell, Alex W. and Taylor, Robert A. and Warner, Jamie H. and Smith, Jason M.},
  title   = {Room-temperature exciton-polaritons with two-dimensional {WS$_2$}},
  journal = {Scientific Reports},
  volume  = {6},
  pages   = {33134},
  year    = {2016},
  doi     = {10.1038/srep33134},
  url     = {https://doi.org/10.1038/srep33134}
}

@article{Kleemann2017,
  author    = {M.‐E. Kleemann and R. Chikkaraddy and E.~M. Alexeev and D. Kos and C. Carnegie and W. Deacon and A. Casalis de Pury and C. Große and B. de Nijs and J. Mertens and A.~I. Tartakovskii and J.~J. Baumberg},
  title     = {Strong‐coupling of {WS}e$_2$ in ultra‐compact plasmonic nanocavities at room temperature},
  journal   = {Nat. Commun.},
  year      = {2017},
  volume    = {8},
  pages     = {1296},
  doi       = {10.1038/s41467-017-01398-3},
  month     = nov,
  day       = {3}
}

@article{Sheinfux2024,
  author  = {Herzig Sheinfux, Hanan and Orsini, Lorenzo and Jung, Minwoo and Torre, Iacopo and Ceccanti, Matteo and Marconi, Simone and Maniyara, Rinu and Barcons Ruiz, David and H{\"o}tger, Alexander and Bertini, Ricardo and Castilla, Sebasti{\'a}n and Hesp, Niels C. H. and Janzen, Eli and Holleitner, Alexander and Pruneri, Valerio and Edgar, James H. and Shvets, Gennady and Koppens, Frank H. L.},
  title   = {High-quality nanocavities through multimodal confinement of hyperbolic polaritons in hexagonal boron nitride},
  journal = {Nature Materials},
  volume  = {23},
  number  = {4},
  pages   = {499--505},
  year    = {2024},
  doi     = {10.1038/s41563-023-01785-w},
  url     = {https://doi.org/10.1038/s41563-023-01785-w}
}

@article{Andolina2026,
  title = {Quantum electrodynamics of graphene Landau levels in a deep-subwavelength hyperbolic phonon-polariton cavity},
  author = {Andolina, Gian Marcello and Ceccanti, Matteo and Turini, Bianca and Riolo, Riccardo and Polini, Marco and Schir\`o, Marco and Koppens, Frank H. L.},
  journal = {Phys. Rev. Res.},
  volume = {8},
  issue = {2},
  pages = {023044},
  numpages = {28},
  year = {2026},
  month = {Apr},
  publisher = {American Physical Society},
  doi = {10.1103/69jg-sq8r},
  url = {https://link.aps.org/doi/10.1103/69jg-sq8r}
}

@article{Chervy2018,
  author  = {Chervy, Thibault and Azzini, Stefano and Lorchat, Etienne and Wang, Shaojun and Gorodetski, Yuri and Hutchison, James A. and Berciaud, St{\'e}phane and Ebbesen, Thomas W. and Genet, Cyriaque},
  title   = {Room Temperature Chiral Coupling of Valley Excitons with Spin-Momentum Locked Surface Plasmons},
  journal = {ACS Photonics},
  volume  = {5},
  number  = {4},
  pages   = {1281--1287},
  year    = {2018},
  doi     = {10.1021/acsphotonics.7b01032},
  url     = {https://doi.org/10.1021/acsphotonics.7b01032}
}

@article{Dirnberger2023,
  author  = {Dirnberger, Florian and Quan, Jiamin and Bushati, Rezlind and Diederich, Geoffrey M. and Florian, Matthias and Klein, Julian and Mosina, Kseniia and Sofer, Zdenek and Xu, Xiaodong and Kamra, Akashdeep and Garc{\'i}a-Vidal, Francisco J. and Al{\`u}, Andrea and Menon, Vinod M.},
  title   = {Magneto-optics in a van der Waals magnet tuned by self-hybridized polaritons},
  journal = {Nature},
  volume  = {620},
  number  = {7974},
  pages   = {533--537},
  year    = {2023},
  doi     = {10.1038/s41586-023-06275-2},
  url     = {https://doi.org/10.1038/s41586-023-06275-2}
}

@article{Kipp2025CavityElectrodynamics,
  author  = {Kipp, Gunda and Bretscher, Hope M. and Schulte, Benedikt and
             Herrmann, Dorothee and Kusyak, Kateryna and
             Day, Matthew W. and Kesavan, Sivasruthi and
             Matsuyama, Toru and Li, Xinyu and
             Langner, Sara Maria and Hagelstein, Jesse and
             Sturm, Felix and Potts, Alexander M. and
             Eckhardt, Christian J. and Huang, Yunfei and
             Watanabe, Kenji and Taniguchi, Takashi and
             Rubio, Angel and Kennes, Dante M. and
             Sentef, Michael A. and Baudin, Emmanuel and
             Meier, Guido and Michael, Marios H. and
             McIver, James W.},
  title   = {Cavity electrodynamics of van der Waals heterostructures},
  journal = {Nature Physics},
  volume  = {21},
  number  = {12},
  pages   = {1926--1933},
  year    = {2025},
  doi     = {10.1038/s41567-025-03064-8},
  url     = {https://doi.org/10.1038/s41567-025-03064-8}
}

@article{Andolina2022NonPerturbative,
  author       = {Gian Marcello Andolina and Francesco M. D. Pellegrino and Alberto Mercurio and Oreste Di Stefano and Marco Polini and Salvatore Savasta},
  title        = {A non-perturbative no-go theorem for photon condensation in approximate models},
  journal      = {Eur. Phys. J. Plus},
  volume       = {137},
  pages        = {1348},
  year         = {2022},
  doi          = {10.1140/epjp/s13360-022-03571-0},
  url          = {https://doi.org/10.1140/epjp/s13360-022-03571-0}
}

@article{Guerci2020,
  author       = {Guerci, Daniele and Simon, Pascal and Mora, Christophe},
  title        = {Superradiant Phase Transition in Electronic Systems and Emergent Topological Phases},
  journal      = {Phys. Rev. Lett.},
  volume       = {125},
  number       = {25},
  pages        = {257604},
  year         = {2020},
  publisher    = {American Physical Society},
  doi          = {10.1103/PhysRevLett.125.257604}
}

@article{Dmytruk2021,
  author       = {Dmytruk, Olesia and Schir\'o, Marco},
  title        = {Gauge fixing for strongly correlated electrons coupled to quantum light},
  journal      = {Phys. Rev. B},
  volume       = {103},
  number       = {7},
  pages        = {075131},
  year         = {2021},
  publisher    = {American Physical Society},
  doi          = {10.1103/PhysRevB.103.075131}
}

@article{Andolina2019,
  author       = {Andolina, G. M. and Pellegrino, F. M. D. and Giovannetti, V. and MacDonald, A. H. and Polini, M.},
  title        = {Cavity quantum electrodynamics of strongly correlated electron systems: A no-go theorem for photon condensation},
  journal      = {Phys. Rev. B},
  volume       = {100},
  number       = {12},
  pages        = {121109(R)},
  year         = {2019},
  publisher    = {American Physical Society},
  doi          = {10.1103/PhysRevB.100.121109}
}

@article{Bloch2022,
  title = {Strongly correlated electron--photon systems},
  author = {Bloch, Jacqueline and Cavalleri, Andrea and Galitski, Victor and Hafezi, Mohammad and Rubio, Angel},
  journal = {Nature},
  volume = {606},
  number = {7912},
  pages = {41--48},
  year = {2022},
  doi = {10.1038/s41586-022-04726-w},
  url = {https://doi.org/10.1038/s41586-022-04726-w}
}

@article{Hubener2024,
  title = {Quantum materials engineering by structured cavity vacuum fluctuations},
  author = {H{\"u}bener, Hannes and Vi{\~n}as Bostr{\"o}m, Emil and Claassen, Martin and Latini, Simone and Rubio, Angel},
  journal = {Mater. Quantum Technol.},
  volume = {4},
  number = {2},
  pages = {023002},
  year = {2024},
  doi = {10.1088/2633-4356/ad4e8b},
  url = {https://doi.org/10.1088/2633-4356/ad4e8b}
}

@article{Schlawin2022,
    author = {Schlawin, F. and Kennes, D. M. and Sentef, M. A.},
    title = {Cavity quantum materials},
    journal = {Appl. Phys. Rev.},
    volume = {9},
    number = {1},
    pages = {011312},
    year = {2022},
    month = {02},
    issn = {1931-9401},
    doi = {10.1063/5.0083825},
    url = {https://doi.org/10.1063/5.0083825},
}


\clearpage


\setcounter{equation}{0}
\setcounter{figure}{0}
\setcounter{table}{0}
\setcounter{page}{1}
\setcounter{section}{0}
\setcounter{subsection}{0}
\setcounter{secnumdepth}{2}

\renewcommand{\thesubsection}{\Alph{subsection}}

\renewcommand{\theequation}{S\arabic{equation}}
\renewcommand{\thefigure}{S\arabic{figure}}
\renewcommand{\thetable}{S\arabic{table}}


\onecolumngrid

\begin{center}

{\large\bfseries
Supplementary Material for\\[0.15cm]
``Electron-Mediated Phonon Polaritons in Quantum Cavities''
}

\vspace{0.35cm}

Mattia Udina$^{1,2}$,
David Hagenm\"uller$^{2}$,
and Indranil Paul$^{1}$

\vspace{0.25cm}

{\small
$^1$\emph{Laboratoire Mat\'eriaux et Ph\'enom\`enes Quantiques,
Universit\'e Paris Cit\'e, CNRS, 75205 Paris, France}
\\
$^2$\emph{Institut de Physique et Chimie des Mat\'eriaux de Strasbourg
(UMR 7504),\\ Universit\'e de Strasbourg and CNRS,
Strasbourg, 67200, France}
}

\end{center}

\vspace{0.35cm}

\begin{center}
\begin{minipage}{0.90\textwidth}
\small
\noindent
This Supplementary Material provides additional details on the microscopic theory and calculations presented in the main text. Section \ref{phpol_pro} derives the gauge-invariant phonon-polariton propagator using a diagrammatic approach. Section \ref{ext_mtl} extends the theory to multilayer geometries and discusses the role of the cavity filling factor. Section \ref{TR_intra} examines the constraints imposed by time-reversal symmetry on the mixed current--phonon response, while Sec.~\ref{micro_sec} presents the microscopic honeycomb-lattice model and the parameters used for NbSe$_2$, hBN, and gapped graphene.
\end{minipage}
\end{center}

\vspace{0.4cm}

\twocolumngrid

\subsection{Derivation of the phonon-polariton propagator}
\label{phpol_pro}


Here we derive the phonon-polariton propagator describing the
coupled dynamics of the phonon and cavity-photon modes mediated by electronic
excitations, using a diagrammatic approach. Within the same approximations
adopted in the main text, we restrict to the diagonal polarization components.
In particular, we first focus on the phononic sector and derive the $(2,2)$ component
of the inverse kernel, $[\hat M(\eta)^{-1}_{\alpha\alpha}]_{22}
=
\tilde B(\eta)_{\alpha\alpha}$, which corresponds to the phonon propagator dressed by its coupling to the
cavity photon. Here $\eta=(i\Omega_n, \mathbf{q})$, where $\Omega_n=2\pi nT$ are the bosonic Matsubara frequencies and $\textbf{q}$ is the in-plane wavevector, and we set $\hbar=1$. The analogous $(1,1)$ component,
$[\hat M(\eta)^{-1}_{\alpha\alpha}]_{11}$, describes the photonic sector,
i.e.\ the cavity photon dressed by its hybridization with the phonon mode.
The full kernel $\hat M(\eta)_{\alpha\alpha}$ can then be reconstructed
\emph{a posteriori} as we detail below.

\begin{figure}[t]
\includegraphics[width=0.48\textwidth]{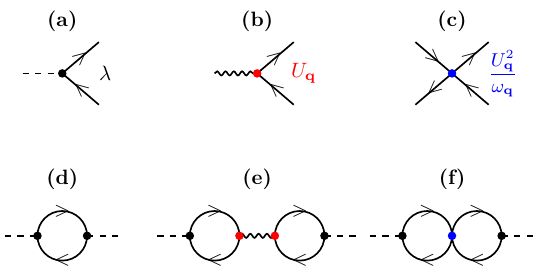}
\caption{Diagrammatic representation of (a) the electron--phonon vertex,
(b) the electron--photon vertex associated with the linear dipole coupling
in Eq.\ (2) of the main text, and (c) the electronic self-interaction arising
from the polarization self-energy in Eq.\ (2). Dashed lines denote phonon propagators, wavy lines photon propagators, and solid lines electronic propagators. (d) Diagrammatic representation of the phonon-phonon correlator $\Pi_{\Gamma\Gamma}$ defined in Eq.\ (\ref{correl}). (e)-(f) Lowest-order self-energy contributions generated by the light-matter interaction.} 
\label{FigS1}
\end{figure}

By focusing first on the phononic sector,
we obtain
\begin{equation}
\tilde B(\eta)^{-1}_{\alpha\alpha}=
B(\eta)^{-1}-
\left[
\lambda^2\Pi_{\Gamma_{\alpha}\Gamma_{\alpha}}(i\Omega_n)
+
\Sigma_{\alpha\alpha}^B(\eta)
\right].
\label{php_pr}
\end{equation}
Here $B(\eta)^{-1}=[(i\Omega_n)^2-\omega_{\rm p,\mathbf{q}}^2]/(2\omega_{\rm p,\mathbf{q}})$ is the inverse bare phonon propagator, with $\omega_{\rm p,\mathbf{q}}$ the bare phonon frequency, while $\lambda$ is the electron-phonon coupling energy. The phonon-phonon correlator
\begin{equation}
\Pi_{\Gamma_{\alpha}\Gamma_{\alpha}}(i\Om_n) \equiv - \frac{1}{N} \int_0^{\be} d \tau
\langle T_{\tau} \hat{\Gamma}_{\alpha}(\tau) \hat{\Gamma}_{\alpha}(0) \rangle e^{i\Om_n \tau},
\label{correl}
\end{equation}
with $N$ the number of lattice sites, is diagrammatically represented in Fig.\ \ref{FigS1}(d) and accounts for the phonon self-energy generated by electronic excitations already in the absence of the cavity field. By contrast, $\Sigma^B$ describes the additional electron-mediated dressing generated by the light--matter interaction. Its lowest-order diagrammatic contributions are shown in Fig.~\ref{FigS1}(e) and Fig.~\ref{FigS1}(f), arising from the linear dipole
coupling and the polarization self-energy, respectively. Taken together, these processes dress the bare photon propagator
$W_{\mathbf q}$ into the gauge-invariant effective propagator
$\widetilde W_{\mathbf q}$ introduced in the main text. Resumming these processes to all orders within the random-phase approximation (RPA), we find
\begin{equation}
\Sigma_{   \alpha\alpha}^B(i\Omega_n,\mathbf{q})= \frac{U_{\mathbf{q}}^2\lambda^2\Pi_{P_{\alpha}\Gamma_{\alpha}}(i\Omega_n) \Pi_{P_{\alpha} \Gamma_{\alpha}}(-i\Omega_n) \widetilde W_{\mathbf{q}}(i\Omega_n)}{1-U_{\mathbf{q}}^2 \Pi_{P_{\alpha} P_{\alpha}}(i\Omega_n)\widetilde W_{\mathbf{q}}(i\Omega_n)},
\label{sigma1}
\end{equation}
where, by definition, $\Pi_{P_{\alpha}\Gamma_{\alpha}}(-i\Omega_n)=\Pi_{\Gamma_{\alpha} P_{\alpha}}(i\Omega_n)$ and $U_{\mathbf{q}}$ is the light-matter interaction strength defined in the main text. The polarization-polarization correlator $\Pi_{P_{\alpha}P_{\alpha}}$ and the mixed correlator $\Pi_{\Gamma_{\alpha} P_{\alpha}}$ are defined analogously to Eq.\ (\ref{correl}), with the electron-phonon operator $\hat\Gamma_{\alpha}=\sum_{\mathbf k,\mu\nu \sigma} f_{\mathbf k}^{\mu\nu}
c^\dagger_{\mathbf k,\mu \sigma}
c_{\mathbf k,\nu \sigma}$ replaced, where appropriate, by the electronic polarization operator $\hat P_{\alpha} =\sum_{\mathbf{R},\mu,\sigma} R_{\alpha} c^{\dagger}_{\mathbf{R}, \mu,\sigma} c_{\mathbf{R}, \mu, \sigma}$. 

Using the Lehmann representation, the polarization-polarization correlator reads
\begin{equation}
\Pi_{P_{\alpha}P_{\alpha}}(i\Omega_n) = -\frac{1}{ZN}\sum_{n,m} [\hat P_{\alpha}]_{nm}[\hat P_{\alpha}]_{mn}
\frac{e^{-\beta E_m}-e^{-\beta E_n}}
{i\Omega_n+E_n-E_m},
\label{PiPP1}
\end{equation}
where $Z=\sum_n e^{-\beta E_n}$ is the partition function, $[\hat P_{\alpha}]_{nm}=\langle n|\hat P_{\alpha}|m\rangle$ denotes the polarization matrix element between the electronic eigenstates $|n\rangle$ and $|m\rangle$, and $\hat{\mathcal{H}}_{\rm e}|n\rangle=E_n|n\rangle$ defines their corresponding energies. In the following the normalization factor
$1/(NZ)$ is left implicit for compactness.  Introducing $E_{nm}\equiv E_n-E_m$, Eq.\ (\ref{PiPP1}) can be rewritten as
\begin{equation}
\begin{aligned}
\Pi_{P_{\alpha}P_{\alpha}}(i\Omega_n)
=&-\frac{1}{(i\Omega_n)^2}
\sum_{n,m}
[\hat P_{\alpha}]_{nm}[\hat P_{\alpha}]_{mn} e^{-\beta E_n}
\Bigg[
2E_{nm}
\\
& +E_{nm}^2
\left(
\frac{1}{i\Omega_n-E_{nm}}
-\frac{1}{i\Omega_n+E_{nm}}
\right)
\Bigg].
\label{PiPP2}
\end{aligned}
\end{equation}
The commutation relations imply $\langle n|[    \hat{\mathcal{H}}_{\rm e},\hat P_{\alpha}]|m\rangle =E_{nm}[\hat P_{\alpha}]_{nm}$. As a consequence, the term linear in $E_{nm}$ in Eq.\ (\ref{PiPP2}) can be expressed as 
\begin{equation}
\begin{aligned}
\langle n|[\hat P_{\alpha},[\hat P_{\alpha},\hat{\mathcal{H}}_{\rm e}]]|n\rangle
=& \sum_m
\Big[ \langle n|\hat P_{\alpha}|m\rangle
\langle m|[\hat P_{\alpha},\hat{\mathcal{H}}_{\rm e}]|n\rangle
\\
&- \langle n|[\hat P_{\alpha},\hat{\mathcal{H}}_{\rm e}]|m\rangle \langle m|\hat P_{\alpha}|n\rangle \Big] \\
=& 2\sum_m
E_{nm} [\hat P_{\alpha}]_{nm}[\hat P_{\alpha}]_{mn},
\end{aligned}
\end{equation}
while the quadratic term becomes
\begin{equation}
E_{nm}^2[\hat P_{\alpha}]_{nm}[\hat P_{\alpha}]_{mn}
=
-\langle n|[\hat P_{\alpha},\hat{\mathcal{H}}_{\rm e}]|m\rangle
\langle m|[\hat P_{\alpha},\hat{\mathcal{H}}_{\rm e}]|n\rangle,
\label{comm2}
\end{equation}
such that Eq.\ (\ref{PiPP2}) takes the form
\begin{equation}
\begin{aligned}
\Pi_{P_{\alpha}P_{\alpha}}(i\Omega_n) = &-\frac{1} {(i\Omega_n)^2}
\Bigg[ \sum_n
\langle n|[\hat P_{\alpha},[\hat P_{\alpha},\hat{\mathcal{H}}_{\rm e}]]|n\rangle
e^{-\beta E_n}
\\
& -\sum_{n,m}
\langle n|[\hat P_{\alpha},\hat{\mathcal{H}}_{\rm e}]|m\rangle
\langle m|[\hat P_{\alpha},\hat{\mathcal{H}}_{\rm e}]|n\rangle
e^{-\beta E_n} \\
& \times
\left(
\frac{1}{i\Omega_n-E_{nm}}
-\frac{1}{i\Omega_n+E_{nm}}
\right)
\Bigg].
\label{PiPP3}
\end{aligned}
\end{equation}

Since the polarization operator is formally ill-defined in a thermodynamically large system, it is more convenient to recast Eq.\ (\ref{PiPP3}) directly in
terms of the current operator \cite{resta}. The connection follows from the Heisenberg
equation, i.e.\
\begin{equation}
\hat j_{\alpha}
\equiv
\frac{\partial \hat P_{\alpha}}{\partial t}
\propto
-i[\hat P_{\alpha},\hat{\mathcal H}_{\rm e}].
\label{currentcomm}
\end{equation}
For the tight-binding Hamiltonian considered here, this relation yields
the Peierls current operator introduced in the main text, i.e.\ $\hat j_\alpha
=
\sum_{\mathbf k,\mu\nu,\sigma}
\partial_{k_\alpha}\epsilon_{\mathbf k}^{\mu\nu}
c^\dagger_{\mathbf k,\mu,\sigma}
c_{\mathbf k,\nu,\sigma}$.

Equation (\ref{PiPP3}) can then be directly related to the full
current-current response. In particular, the first term corresponds to
the diamagnetic contribution,
\begin{equation}
\chi^{\mathrm{dia}}_{j_\alpha j_\alpha}
\equiv
-\sum_n
\langle n|[\hat P_{\alpha},\hat j_{\alpha}]|n\rangle
e^{-\beta E_n},
\label{chidia}
\end{equation}
where $[\hat P_{\alpha},\hat j_{\alpha}]
=
-\sum_{\mathbf k,\mu\nu,\sigma}
\partial^2_{k_\alpha}\epsilon_{\mathbf k}^{\mu\nu}
c^\dagger_{\mathbf k,\mu,\sigma}
c_{\mathbf k,\nu,\sigma}$.
The second term in Eq.~(\ref{PiPP3}), instead, becomes the paramagnetic
current-current correlator. Using Eq.~(\ref{currentcomm}), it can be written
in Lehmann representation as
\begin{equation}
\begin{aligned}
\chi_{j_{\alpha}j_{\alpha}}(i\Omega_n)
={}&
\sum_{n,m}
[\hat j_{\alpha}]_{nm}
[\hat j_{\alpha}]_{mn}
e^{-\beta E_n}
\\
&\times
\left(
\frac{1}{i\Omega_n-E_{nm}}
-
\frac{1}{i\Omega_n+E_{nm}}
\right).
\end{aligned}
\label{chijj}
\end{equation}
The polarization-polarization response can then be written as
\begin{align}
\Pi_{P_{\alpha}P_{\alpha}}(i\Omega_n)
&=
\frac{1}{(i\Omega_n)^2}
\left[
\chi^{\mathrm{dia}}_{j_\alpha j_\alpha}
+
\chi_{j_{\alpha}j_{\alpha}}(i\Omega_n)
\right] \nonumber \\
&=
\frac{\Pi_{j_{\alpha}j_{\alpha}}(i\Omega_n)}
{(i\Omega_n)^2},
\label{PiPP_current}
\end{align}
where $\Pi_{j_{\alpha}j_{\alpha}}$ denotes the full gauge-invariant
current-current response, including both diamagnetic and paramagnetic contributions. Gauge invariance requires that, in the static and uniform limit, a constant vector potential cannot induce a physical current, implying $\Pi_{j_{\alpha}j_{\alpha}}(0)=0$ \cite{Li2020,Dmytruk2021, Andolina2019}. Consequently, the diamagnetic contribution exactly cancels the static paramagnetic response, $\chi^{\text{dia}}_{j_{\alpha}j_{\alpha}}=-\chi_{j_{\alpha}j_{\alpha}}(0)$. After analytical continuation to real frequencies, $i\Omega_n\to\omega+i0^+$, the full current-current response can then be expressed as
\begin{equation}
\Pi_{j_{\alpha} j_{\alpha}}(\om) = \chi_{j_{\alpha} j_{\alpha}}(\om)  - \chi_{j_{\alpha} j_{\alpha}}(0).
\end{equation}

Focusing on the mixed correlator $\Pi_{P_{\alpha}\Gamma_{\alpha}}$, and proceeding in close analogy with Eqs.\ (\ref{PiPP1}) and (\ref{PiPP2}), we obtain
\begin{equation}
\begin{aligned}
\Pi_{P_{\alpha}\Gamma_{\alpha}}(i\Omega_n)
=&-\frac{1}{i\Omega_n}
\sum_{n,m}
[\hat P_{\alpha}]_{nm}[\hat \Gamma_{\alpha}]_{mn} \left (e^{-\beta E_m}-e^{-\beta E_n} \right)
\\
& \times \left( 1-
\frac{E_{nm}}{i\Omega_n+E_{nm}}
\right).
\label{PiPG1}
\end{aligned}
\end{equation}
Once noted that, in analogy with Eq.\ (\ref{currentcomm}), one finds
\begin{equation}
[\hat P_{\alpha}, \hat \Gamma_{\alpha}]=i \sum_{\mathbf k,\mu\nu,\sigma} \partial_ {k_\alpha} f_{\mathbf k}^{\mu\nu}
c^\dagger_{\mathbf k,\mu,\sigma}
c_{\mathbf k,\nu,\sigma},
\end{equation}
we can identify a diamagnetic-like contribution to the mixed current-phonon response, differing from the
usual diamagnetic term by an overall factor $i$, i.e.\ 
\begin{equation}
\begin{aligned}
i\chi^{\text{dia}}_{j_\alpha \Gamma_\alpha}
&\equiv
-\sum_{n,m}
[\hat P_{\alpha}]_{nm}[\hat \Gamma_{\alpha}]_{mn} \left (e^{-\beta E_m}-e^{-\beta E_n} \right)\\
&=\sum_{n}
\langle n|[\hat P_{\alpha},\hat \Gamma_{\alpha}]|n\rangle
e^{-\beta E_n}.
\end{aligned}
\end{equation}
The second term in Eq.\ (\ref{PiPG1}) can instead be rewritten in terms of the current operator as
\begin{equation}
\begin{aligned}
i\chi_{j_\alpha \Gamma_\alpha}(i\Omega_n)
&\equiv
-i\sum_{n,m}
[\hat j_{\alpha}]_{nm}[\hat \Gamma_{\alpha}]_{mn} \frac{\left(e^{-\beta E_m}-e^{-\beta E_n} \right)}{i\Omega_n+E_{nm}}.
\end{aligned}
\label{chiJGpara}
\end{equation}
Combining the two contributions, the mixed
polarization-phonon correlator can therefore be written
as
\begin{align}
\Pi_{P_{\alpha}\Gamma_{\alpha}}(i\Omega_n) &=
\frac{i}{i\Omega_n}
\left[ \chi^{\rm dia}_{j_\alpha \Gamma_\alpha} + \chi_{j_{\alpha}\Gamma_{\alpha}
}(i\Omega_n)
\right]\nonumber \\
&=\frac{i\Pi_{j_{\alpha}\Gamma_{\alpha}}(i\Omega_n)}
{i\Omega_n},
\end{align}
where $\Pi_{j_{\alpha}\Gamma_{\alpha}}(i\Omega_n)$ denotes the full, gauge-invariant mixed current-phonon response. By the same reasoning, reversing the operator order gives $\Pi_{\Gamma_{\alpha}P_{\alpha}}(i\Omega_n)=-i\Pi_{\Gamma_{\alpha}j_{\alpha}}(i\Omega_n)/(i\Omega_n)$, with $\Pi_{\Gamma_{\alpha}j_{\alpha}}(i\Omega_n)=\Pi_{j_{\alpha}\Gamma_{\alpha}}(-i\Omega_n)$ by construction.

The appearance of the diamagnetic-like term is essential for preserving gauge invariance, in complete analogy with the current-current response. Indeed, using $[j_{\alpha}]_{nm}=iE_{nm}[P_{\alpha}]_{nm}$, the static limit of Eq.\ (\ref{chiJGpara}) yields
\begin{equation}
i\chi_{j_{\alpha}\Gamma_{\alpha}}(0)= -i\chi^{\rm dia}_{j_{\alpha}\Gamma_{\alpha}}.
\end{equation}
Therefore, after analytical continuation to real frequencies the full gauge-invariant mixed response can be written as
\begin{equation}
\Pi_{j_{\alpha}\Gamma_{\alpha}}(\omega)=\chi_{j_{\alpha}\Gamma_{\alpha}}(\omega)-\chi_{j_{\alpha}\Gamma_{\alpha}}(0),
\end{equation}
which explicitly ensures $\Pi_{j_{\alpha}\Gamma_{\alpha}}(0)=0$.

Using the current-current and current-phonon correlators introduced above, the cavity-induced phonon self-energy can then be rewritten as
\begin{equation}
\Sigma_{\alpha\alpha}^B(\omega,\mathbf{q})=
\frac{\left(\frac{2U_{\mathbf{q}}^2\lambda^2}{\omega_{\rm c,\mathbf{q}}} \right)
\Pi_{j_{\alpha}\Gamma_{\alpha}}(\omega)
\Pi_{j_{\alpha}\Gamma_{\alpha}}(-\omega)}
{\omega^2-\omega_{\rm c,\mathbf{q}}^2-
\left(\frac{2U_{\mathbf{q}}^2}{\omega_{\rm c,\mathbf{q}}} \right)
\Pi_{j_{\alpha}j_{\alpha}}(\omega)}.
\label{sigma2}
\end{equation}

To make the connection with the matrix formulation introduced in the main text explicit,
we note that the dressed phonon propagator can be obtained as the $(2,2)$ component
of the inverse of a generic $2\times2$ kernel. Indeed, for
\begin{equation}
\hat M=
\begin{pmatrix}
A & C\\
D & F
\end{pmatrix},
\end{equation}
one has $[\hat M^{-1}]_{22}
= \left(F-D A^{-1} C\right)^{-1}$. In the present case, the electron-dressed inverse photon propagator is
\begin{equation}
A(\omega,\mathbf q)
=
W_{\mathbf q,\omega}^{-1}
-g_{\mathbf q}^2\Pi_{j_\alpha j_\alpha}(\omega),
\end{equation}
with $W_{\mathbf q,\omega}^{-1}
= \frac{\omega^2-\omega_{\rm c,\mathbf q}^2} {2\omega_{\rm c,\mathbf q}}$, and $g_{\mathbf q}
= \frac{U_{\mathbf q}}{\omega_{\rm c,\mathbf q}}$. The off-diagonal photon-phonon blocks are instead
$ C(\omega, \mathbf{q})
=
i g_{\mathbf q}\lambda
\Pi_{j_\alpha\Gamma_\alpha}(\omega)$ and 
$D(\omega,\mathbf{q})
= -i g_{\mathbf q}\lambda
\Pi_{j_\alpha\Gamma_\alpha}(-\omega)$, such that
\begin{equation}
\begin{aligned}
D A^{-1}C
&=
g_{\mathbf q}^2\lambda^2
\Pi_{j_\alpha\Gamma_\alpha}(-\omega)
A^{-1}(\omega,\mathbf q)
\Pi_{j_\alpha\Gamma_\alpha}(\omega)
\\
&=
\Sigma_{\alpha\alpha}^B(\omega,\mathbf q),
\end{aligned}
\end{equation}
which reproduces Eq.\ (\ref{sigma2}). Thus, identifying the phononic block as $F(\omega,\mathbf q) = B_{\mathbf q,\omega}^{-1}
 -\lambda^2\Pi_{\Gamma_\alpha\Gamma_\alpha}(\omega)$,
the phonon-polariton propagator reads
\begin{equation}
[\hat M^{-1}]_{22}
=
\left[
B_{\mathbf q,\omega}^{-1}
-\lambda^2\Pi_{\Gamma_\alpha\Gamma_\alpha}(\omega)
-\Sigma_{\alpha\alpha}^B(\omega,\mathbf q)
\right]^{-1},
\end{equation}
in agreement with Eq.\ (\ref{php_pr}). The corresponding inverse kernel is
therefore
\begin{equation}
\hat M(\eta)_{\alpha\alpha}=
\begin{pmatrix}
 W_{\mathbf{q}, \omega}^{-1}-g_{\mathbf{q}}^2
 \Pi_{j_{\alpha}j_{\alpha}}(\omega) &
 ig_{\mathbf{q}}\lambda
 \Pi_{j_{\alpha}\Gamma_{\alpha}}(\omega) \\
-ig_{\mathbf{q}}\lambda
 \Pi_{j_{\alpha}\Gamma_{\alpha}}(-\omega) &
 B_{\mathbf q,\omega}^{-1}
-\lambda^2\Pi_{\Gamma_{\alpha}\Gamma_{\alpha}}(\omega)
\end{pmatrix}.
\label{2x2SM}
\end{equation}

The same kernel also provides direct access to the photonic sector. In
particular, the $(1,1)$ component of the inverse matrix defines the
phonon-polariton propagator projected onto the photon channel, i.e.\ the
cavity photon dressed by its hybridization with the phonon mode,
$\tilde A^{-1}=[\hat M^{-1}]_{11}$, with
\begin{equation}
[\hat M^{-1}]_{11}
=
\left[
W^{-1}_{\mathbf q,\omega}
-
g_{\mathbf q}^{\,2}\Pi_{j_\alpha j_\alpha}(\omega)
-
\Sigma^A_{\alpha\alpha}(\omega,\mathbf q)
\right]^{-1},
\end{equation}
where
\begin{equation}
\Sigma^A_{\alpha\alpha}(\omega,\mathbf q)
=
\frac{
2\omega_{\rm p,\mathbf q}\,
g_{\mathbf q}^{\,2}\lambda^2
\Pi_{j_\alpha\Gamma_\alpha}(\omega)
\Pi_{j_\alpha\Gamma_\alpha}(-\omega)
}{
\omega^2-\omega_{\rm p,\mathbf q}^2
-
2\omega_{\rm p,\mathbf q}\lambda^2
\Pi_{\Gamma_\alpha\Gamma_\alpha}(\omega)
}.
\end{equation}
The poles of $[\hat M^{-1}]_{11}$ coincide with those of
$[\hat M^{-1}]_{22}$, since the two propagators describe the same
phonon-polariton eigenmodes projected onto the photonic and phononic
sectors, respectively. The spectral weights in the two channels instead
redistribute between the same polariton branches, reflecting their varying
photonic and phononic character \cite{sellati2025}.

\subsection{Extension to a multilayer geometry}
\label{ext_mtl}

We now generalize the two-dimensional formulation to a film composed of
$n_z$ equivalent layers. We therefore introduce a layer index
$\ell=1,\dots,n_z$ for both the electronic and phononic degrees of freedom. Since the cavity field is assumed to be uniform along the out-of-plane direction, we retain
the uniform phonon mode, corresponding to the normalized combination
\begin{equation}
b_{\mathbf q,\alpha}^{(0)}
=
\frac{1}{\sqrt{n_z}}
\sum_{\ell=1}^{n_z}
b_{\mathbf q,\alpha,\ell}.
\label{b0}
\end{equation}
Projecting the layer-resolved electron--phonon interaction onto this uniform
mode yields
\begin{equation}
\mathcal H_{ep} = \frac{\lambda}{\sqrt{N n_z}}
\sum_{\mathbf q,\alpha,\ell}
\left( b_{\mathbf q,\alpha}^{(0)} + b_{-\mathbf q,\alpha}^{\dagger(0)} \right) \Gamma_{\mathbf q,\alpha,\ell}.
\label{Hep0}
\end{equation}
Since the electronic layers are assumed to be independent, the phonon-phonon response is diagonal in the layer indices. The product of the two electron--phonon vertices therefore carries a factor
$1/(Nn_z)$, which is compensated by the momentum and layer sums in the electronic bubble, $\sum_{\mathbf k,\ell}\sim Nn_z$. As a consequence, the
intrinsic self-energy of the uniform phonon mode remains
independent of the number of layers.

The cavity photon, on the other hand, is a single electromagnetic mode
and couples coherently to the total electronic polarization,
$\mathbf P_{\mathbf q}^{\rm tot}=\sum_\ell \mathbf P_{\mathbf q,\ell}$.
Consequently, the mixed current--phonon and current--current responses
scale as $\sqrt{n_z}$ and $n_z$, respectively. Equivalently, keeping the
single-layer response functions, these collective enhancements can be
accounted for by the rescaling
\begin{equation}
U_{\mathbf q}
\rightarrow
\sqrt{n_z}\,U_{\mathbf q}
= U_{\mathbf q}\sqrt{\frac{d}{a_z}},
\label{UZ}
\end{equation}
with $d=n_z a_z$ the sample thickness and $a_z$ the inter-layer spacing.

We next consider the effect of the finite sample thickness on the
light-matter interaction and, in particular, on the dispersion of the
electron-dressed cavity photon. Eq.\ (\ref{2x2SM}) shows that
$\Pi_{jj}$ accounts for the dielectric screening of the electromagnetic
field by the electronic medium. The current-current response is related to
the two-dimensional optical conductivity through $\Pi_{jj}(\omega)=-i\omega\sigma(\omega)$.
The corresponding bulk dielectric function can then be introduced as
\begin{equation}
\Pi_{jj}(\omega)
=
\frac{\epsilon_0 a_z\omega^2}{e^2}
\left[1-\epsilon(\omega)\right].
\end{equation}
For a sample of thickness $d$ embedded in a cavity of length $L$, the
electron-dressed photon propagator then reads
\begin{equation}
[\hat M_{11}(\mathbf q,\omega)]^{-1}
=
\frac{2\omega_{c,\mathbf q}}
{\epsilon_c(\omega)\omega^2-\omega_{c,\mathbf q}^2},
\end{equation}
with $\epsilon_c(\omega)
=1+\left[\epsilon(\omega)-1\right]d/L$ the cavity dielectric function. This expression provides the macroscopic counterpart of the multilayer enhancement discussed in Eq.\ (\ref{UZ}).
The factor $d/L$ represents the cavity filling fraction and quantifies the
geometric reduction of the electronic screening when the material occupies
only a fraction of the cavity volume. In the fully filled limit ($d=L$), one has
$\epsilon_c(\omega)=\epsilon(\omega)$ and the filling-factor
dependence disappears. In this sense, the photon experiences the same
dielectric environment as in the bulk material, and there is no additional
geometric reduction of the electronic screening associated with the
cavity filling fraction \cite{BarraBurillo2021}. The electromagnetic field is still quantized by the cavity boundary conditions but the dielectric response is that of
the multi-layer medium.

\subsection{Time-reversal symmetry and the absence of intraband contributions to the mixed current-phonon response}
\label{TR_intra}

Here we show that time-reversal symmetry
forbids intraband contributions to the paramagnetic mixed current-phonon
response in the uniform limit ($q\to0$). We consider a generic electronic Hamiltonian 
$\mathcal H_{\rm e}=\sum_{\mathbf k,\sigma}
\Psi^\dagger_{\mathbf k,\sigma}
\hat H_{\rm e,\mathbf k,\sigma}
\Psi_{\mathbf k,\sigma}$,
with
$\Psi_{\mathbf k,\sigma}
=(c_{\mathbf k,\mu,\sigma},c_{\mathbf k,\nu,\sigma})^T$,
and we denote by $\hat j_{\mathbf k,\sigma}$ and
$\hat\Gamma_{\mathbf k,\sigma}$ the current and electron-phonon vertices,
respectively. Neglecting cartesian indices for compactness, the corresponding paramagnetic mixed response is
\begin{equation}
\chi_{j\Gamma}(i\Omega_n)
=
\frac{T}{N}
\sum_{\mathbf k,\omega_n,\sigma}
\mathrm{Tr}\!\left[
\hat j_{\mathbf k,\sigma}
\hat G_{\mathbf k,\sigma}(i\omega_n')
\hat\Gamma_{\mathbf k,\sigma}
\hat G_{\mathbf k,\sigma}(i\omega_n)
\right],
\end{equation}
where $i\omega_n'=i\omega_n+i\Omega_n$ and $\hat G_{\mathbf k,\sigma}(i\omega_n) =
(
i\omega_n-\hat H_{\rm e,\mathbf k,\sigma}
)^{-1}$ is the electronic Green's function. Introducing the band basis,
$\hat H_{\rm e,\mathbf k,\sigma}\vert a,\mathbf k,\sigma\rangle
=
E_{a\sigma}(\mathbf k)\vert a,\mathbf k,\sigma\rangle$, with $E_{a\sigma}(\mathbf{k})$ the dispersion of the band $a$, the Green's
function is diagonal in band space, with $G_{a\sigma}(\mathbf k,i\omega_n) = (i\omega_n-E_{a\sigma}(\mathbf k))^{-1}$ its diagonal components. The response can then be decomposed into intra- and interband contributions.
The intraband part, obtained by retaining terms with the same band index
at the two vertices, reads
\begin{equation}
\begin{aligned}
\chi_{j\Gamma}^{\mathrm{intra}}(i\Omega_n)
=&
\frac{T}{N}
\sum_{\mathbf{k},\omega_n,a,\sigma}
[j_{\mathbf{k},\sigma}]_{aa}
[\Gamma_{\mathbf{k},\sigma}]_{aa}
\\
&\times
G_{a\sigma}(\mathbf{k},i\omega_n')
G_{a\sigma}(\mathbf{k},i\omega_n),
\label{Pi_jGamma_intra}
\end{aligned}
\end{equation}
where
$[j_{\mathbf{k},\sigma}]_{aa}
=
\langle a,\mathbf k,\sigma|
\hat j_{\mathbf k,\sigma}
|a,\mathbf k,\sigma\rangle$,
and analogously for
$[\Gamma_{\mathbf{k},\sigma}]_{aa}$.

For a time-reversal-symmetric system, the electron-phonon vertex
associated with a lattice displacement is even under time reversal,
whereas the current operator is odd. Their diagonal matrix elements
therefore satisfy
\begin{equation}
[\Gamma_{-\mathbf{k},-\sigma}]_{aa}
=
[\Gamma_{\mathbf{k},\sigma}]_{aa},
\qquad
[j_{-\mathbf{k},-\sigma}]_{aa}
=
-[j_{\mathbf{k},\sigma}]_{aa}.
\label{TR_vertices}
\end{equation}
Time-reversal symmetry also implies \cite{morimoto}
\begin{equation}
G_{a,-\sigma}(-\mathbf{k},i\omega_n)
=
G_{a,\sigma}(\mathbf{k},i\omega_n).
\label{TR_Green}
\end{equation}
It follows that the integrand in Eq.~(\ref{Pi_jGamma_intra}) is odd under
the transformation
$(\mathbf{k},\sigma)\rightarrow(-\mathbf{k},-\sigma)$. Therefore, the contributions from the time-reversed states
$(\mathbf{k},\sigma)$ and $(-\mathbf{k},-\sigma)$ cancel identically upon
summation over momentum and spin, leading to $\chi_{j\Gamma}^{\mathrm{intra}}(i\Omega_n)=0$. This cancellation remains valid in the presence of the impurity scattering considered here.

Importantly, this property follows from time-reversal symmetry alone
and does not require inversion symmetry. If inversion symmetry is also
present, however, the electron-phonon vertex can be subject to an
additional constraint. For an infrared-active phonon, the phonon displacement is odd under inversion. Since the electron-phonon Hamiltonian
must remain invariant, the corresponding electronic vertex is inversion
odd. For a spin-independent coupling this implies $[\Gamma_{-\mathbf{k},\sigma}]_{aa} =
-[\Gamma_{\mathbf{k},\sigma}]_{aa}$. Combining this relation with time-reversal symmetry,
$ [\Gamma_{-\mathbf{k},-\sigma}]_{aa} =
[\Gamma_{\mathbf{k},\sigma}]_{aa}$, and assuming spin-degenerate bands, one obtains the stronger condition $[\Gamma_{\mathbf{k},\sigma}]_{aa}=0$, leading to the direct cancellation of $\chi_{j\Gamma}^{\text{intra}}(i\Omega_n)$ for every $\mathbf{k}$. In the noncentrosymmetric case, inversion symmetry does
not impose this additional constraint and
$(\Gamma_{\mathbf{k},\sigma})_{aa}$ can in general be finite.
Nevertheless, time-reversal symmetry still constrains it to be even under
$(\mathbf{k},\sigma)\rightarrow(-\mathbf{k},-\sigma)$, while the diagonal
current matrix element is odd. Hence their product is odd, and the
intraband contribution to $\chi_{j\Gamma}(i\Omega_n)$ vanishes after summation
over the Brillouin zone and spin.

The above argument applies specifically to the intraband contribution to
the paramagnetic mixed response. The latter can nevertheless contain a
finite interband contribution,
$\chi_{j\Gamma}(i\Omega_n)=\chi_{j\Gamma}^{\mathrm{inter}}(i\Omega_n)$,
which is not forbidden by time-reversal symmetry. In particular, starting from the general
Lehmann representation Eq.\ (\ref{chiJGpara}) and performing the analytical
continuation to real frequencies $i\Omega_n\to\omega+i0^+$, one obtains
\begin{equation}
\chi_{j\Gamma}^{\mathrm{inter}}(\omega)
=
-\sum_{\substack{a\neq b,\mathbf{k},\sigma}}
[j_{\mathbf{k},\sigma}]_{ab}
[\Gamma_{\mathbf{k},\sigma}]_{ba}
\frac{e^{-\beta E_b}-e^{-\beta E_a}}
{\omega+i0^++E_{ab}},
\label{chiJGinter}
\end{equation}
with ($a,b$) band indices. For the interband matrix elements, time-reversal symmetry implies
\begin{equation}
[j_{-\mathbf{k},-\sigma}]_{ab}
=
-[j_{\mathbf{k},\sigma}]_{ab}^{*},
\qquad
[\Gamma_{-\mathbf{k},-\sigma}]_{ba}
=
[\Gamma_{\mathbf{k},\sigma}]_{ba}^{*}.
\label{TRSinter}
\end{equation}
Since the band energies are invariant under time reversal, the remaining
factor in Eq.~(\ref{chiJGinter}) is unchanged under
$(\mathbf{k},\sigma)\rightarrow(-\mathbf{k},-\sigma)$. In the
low-frequency off-resonant regime, where no real interband transitions are
excited, the energy
denominator is real. Using Eq.\ (\ref{TRSinter}) one then finds $\chi_{j\Gamma}^{\mathrm{inter}}(\omega)
= - \left[
\chi_{j\Gamma}^{\mathrm{inter}}(\omega)
\right]^*$,
showing that the low-frequency interband mixed response is purely
imaginary. At leading order, its low-frequency behavior can then be written as
\begin{equation}
\lim_{\omega \to 0} \Pi_{j \Gamma}^{\text{inter}}(\omega) = i \omega Z_0,
\end{equation}
where the pre-factor $Z_0$ is directly related to the static Born effective charge of the phonon \cite{stengel22, binci21, sellati2025}.

\subsection{Microscopic honeycomb-lattice model}
\label{micro_sec}

\begin{figure}[t]
\includegraphics[width=0.48\textwidth]{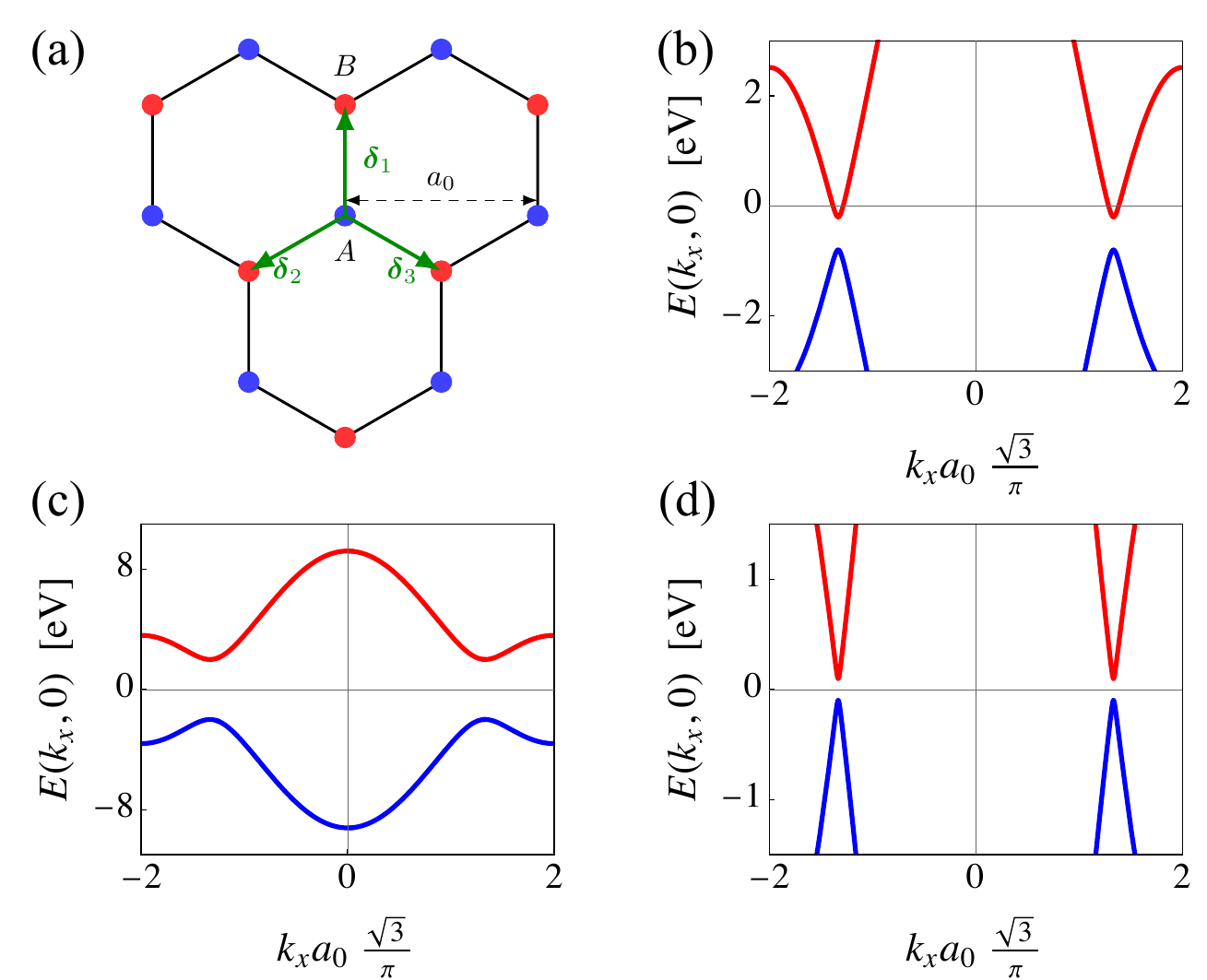}
\caption{(a) Bipartite honeycomb lattice used in the model, with inequivalent sublattices $A$ and $B$ and nearest-neighbor vectors $\boldsymbol{\delta}_{1,2,3}$ defined in the text. (b)-(d) Corresponding electronic band structure along $k_y=0$ for the parameter sets used to describe, respectively, the metallic regime representative of NbSe$_2$, the wide-gap insulating regime representative of hBN, and semiconducting gapped graphene. Red and blue lines denote the upper and lower bands, respectively, and the Fermi level is set to zero energy.} 
\label{FigS2}
\end{figure}

As a specific model, we consider spinless electrons on a honeycomb lattice with two inequivalent
sites ($A$, $B$) per unit cell. Retaining only nearest-neighbor hopping,
the real-space Hamiltonian reads
\begin{equation}
\begin{aligned}
H_e=
&-t\sum_{\mathbf R,j}
c^\dagger_{A,\mathbf R}
c_{B,\mathbf R+\boldsymbol\delta_j}
+\mathrm{h.c.}\\
&+\Delta\sum_{\mathbf R}
\left(n_{A,\mathbf R}-n_{B,\mathbf R}\right)
-\mu\sum_{\mathbf R,s}n_{s,\mathbf R},
\end{aligned}
\end{equation}
with $\boldsymbol\delta_j$ the three nearest-neighbor vectors. As a convention, we choose $\boldsymbol\delta_1 =
a_0\left(0,\frac{1}{\sqrt3}\right)$, $
\boldsymbol\delta_2=
a_0\left(-\frac12,-\frac{1}{2\sqrt3}\right)$ and $ \boldsymbol\delta_3=
a_0\left(\frac12,-\frac{1}{2\sqrt3}\right)$, see Fig.\ \ref{FigS2}(a). 
Introducing the spinor
$\Psi_{\mathbf k}=(c_{A,\mathbf k},c_{B,\mathbf k})^T$, the Hamiltonian can
be written as $\mathcal{H}_{\rm e}=\sum_{\mathbf k}
\Psi^\dagger_{\mathbf k}\hat H_{\rm e,\mathbf k}\Psi_{\mathbf k}$,
with
\begin{equation}
\hat H_{e,\mathbf k}=
\begin{pmatrix}
\Delta-\mu & -t\epsilon_{\mathbf k}\\
-t\epsilon_{\mathbf k}^* &-\Delta-\mu
\end{pmatrix},
\end{equation}
and $\epsilon_{\mathbf k} =\sum_{j=1}^3 e^{i\mathbf k\cdot\boldsymbol\delta_j}=e^{i\frac{a_0k_y}{\sqrt{3}}}+e^{ia_0\left( -\frac{k_x}{2}-\frac{k_y}{2\sqrt{3}}\right)}+e^{ia_0\left(\frac{k_x}{2}-\frac{k_y}{2\sqrt{3}}  \right)}$. The electronic band dispersions are obtained by diagonalizing $\hat H_{\rm e,\mathbf{k}}$, leading to $E_{a,b}(\mathbf{k})=\pm \sqrt{t^2|\epsilon_{\mathbf{k}}|^2+\Delta^2}-\mu$. The (paramagnetic) Peierls current vertex is given by
\begin{equation}
\hat j_{\mathbf{k},\alpha}= \frac{\partial \hat H_{\rm e,\mathbf{k}}}{\partial k_{\alpha}}= \begin{pmatrix} 0 & -t\partial_{k_{\alpha}} \epsilon_{\mathbf{k}} \\ -t\partial_{k_{\alpha}} \epsilon_{\mathbf{k}}^* &0 \end{pmatrix},
\end{equation}
with $\alpha=(x,y)$ the Cartesian index. We next introduce the doubly degenerate zone-center in-plane TO phonons. For these modes the two sublattices undergo opposite displacements. We define the relative displacement $\mathbf u\equiv \mathbf u_B-\mathbf u_A$,
and introduce the phonon normal coordinate $Q_\alpha$ through
$u_\alpha \propto Q_\alpha$. The displacement changes the equilibrium length of the $j$-th
nearest-neighbor bond $r_0$ by $\delta r_j
\simeq
\hat{\boldsymbol\delta}_j\cdot\mathbf u $, where
$\hat{\boldsymbol\delta}_j=\boldsymbol\delta_j/|\boldsymbol\delta_j|$.
Consequently, to linear order in the lattice displacement, the hopping
amplitude becomes
\begin{equation}
t_j=t+\delta t_j,
\qquad
\delta t_j
= \left.\frac{\partial t}{\partial r}\right|_{r=r_0}
\hat{\boldsymbol\delta}_j\cdot\mathbf u.
\end{equation}
For a phonon polarized along $\alpha$ direction,
$\mathbf u=u_\alpha\hat{\mathbf e}_\alpha$, and therefore
\begin{equation}
\delta t_j \propto Q_{\alpha} \hat{\delta}_{j,\alpha},
\end{equation}
where
$\hat{\delta}_{j,\alpha}
=\hat{\boldsymbol\delta}_j\cdot\hat{\mathbf e}_\alpha$. The corresponding electron-phonon interaction then reads
\begin{equation}
\mathcal H_{\rm ep}
=
\frac{\lambda}{\sqrt{N}}\sum_{\mathbf k,\alpha}
Q_\alpha\,
\Psi^\dagger_{\mathbf k}
\hat\Gamma_{\mathbf k,\alpha}
\Psi_{\mathbf k},
\end{equation}
where all momentum-independent prefactors have been absorbed into the
coupling constant $\lambda$. The vertex matrix is given by \cite{fachin24,guinea07}
\begin{equation}
\hat\Gamma_{\mathbf k,\alpha}
=
\begin{pmatrix}
0 &
-\sum_j
\hat\delta_{j,\alpha}
e^{i\mathbf k\cdot\boldsymbol\delta_j}
\\
-\sum_j
\hat \delta_{j,\alpha}
e^{-i\mathbf k\cdot\boldsymbol\delta_j}
&0
\end{pmatrix}.
\end{equation}
Using $\partial_{k_\alpha}\epsilon_{\mathbf k} = ir_0\sum_j \hat\delta_{j,\alpha}e^{i\mathbf k\cdot\boldsymbol\delta_j}$ and absorbing $1/r_0$ into $\lambda$,
the vertex can then be expressed as
\begin{equation}
\hat\Gamma_{\mathbf k,\alpha}
=
\begin{pmatrix}
0 & i\partial_{k_\alpha}\epsilon_{\mathbf k}\\
 -i\partial_{k_\alpha}\epsilon_{\mathbf k}^* &0
\end{pmatrix}.
\end{equation}

We now have all the ingredients required to evaluate the electronic
correlators entering the phonon-polariton propagator in
Eq.\ (\ref{php_pr}). Focusing on the $\alpha=x$ mode, we suppress Cartesian
indices for compactness and define
\begin{equation}
\Pi_{\Gamma\Gamma}(i\Omega_n)=
\frac{T}{N}
\sum_{\mathbf{k},\omega_n}
\mathrm{Tr}\!\left[
\hat\Gamma_{\mathbf{k}}
\hat G_{\mathbf{k}}(i\omega_n')
\hat\Gamma_{\mathbf{k}}
\hat G_{\mathbf{k}}(i\omega_n)
\right],
\end{equation}
for the phonon-phonon correlator, 
\begin{equation}
\begin{aligned}
\chi_{jj}(i\Omega_n)
&=
\frac{T}{N}
\sum_{\mathbf{k},\omega_n}
\mathrm{Tr}\!\left[
\hat j_{\mathbf{k}}
\hat G_{\mathbf{k}}(i\omega_n')
\hat j_{\mathbf{k}}
\hat G_{\mathbf{k}}(i\omega_n)
\right],\\
\chi_{jj}^{\mathrm{dia}}
&=
\frac{T}{N}
\sum_{\mathbf{k},\omega_n}
\mathrm{Tr}\!\left[
\partial_{k_x}\hat j_{\mathbf{k}}
\hat G_{\mathbf{k}}(i\omega_n)
\right],
\end{aligned}
\end{equation}
for the current-current response, and 
\begin{equation}
\begin{aligned}
\chi_{j\Gamma}(i\Omega_n)
&=
\frac{T}{N}
\sum_{\mathbf{k},\omega_n}
\mathrm{Tr}\!\left[
\hat j_{\mathbf{k}}
\hat G_{\mathbf{k}}(i\omega_n')
\hat\Gamma_{\mathbf{k}}
\hat G_{\mathbf{k}}(i\omega_n)
\right],\\
\chi_{j\Gamma}^{\mathrm{dia}}
&=
\frac{T}{N}
\sum_{\mathbf{k},\omega_n}
\mathrm{Tr}\!\left[
\partial_{k_x}\hat\Gamma_{\mathbf{k}}
\hat G_{\mathbf{k}}(i\omega_n)
\right],
\label{bubble}
\end{aligned}
\end{equation}
for the mixed current-phonon response. Here $i\omega_n'\equiv i\omega_n+i\Omega_n$ and $\hat G_{\mathbf{k}}(i\omega_n) = 
(i\omega_n-\hat H_{\rm e,\mathbf{k}})^{-1}$ is the electronic Green's function. For the sake of explicit calculations, it is convenient to work in the band basis where the Green's function is diagonal
\begin{equation}
\hat G_{\mathbf{k}}(i\omega_n)=\hat U_{\mathbf{k}}\begin{pmatrix} G_a(\mathbf{k},i\omega_n) & 0 \\ 0 & G_b(\mathbf{k},i\omega_n) \end{pmatrix}\hat U_{\mathbf{k}}^{\dagger},
\end{equation}
with $G_{a,b}(\mathbf{k},i\omega_n)=[i\omega_n- E_{a,b}(\mathbf{k})]^{-1}$ and $\hat U_{\mathbf{k}}$ the unitary transformation from the orbital to the band basis. As detailed in Ref.\ \cite{udina26}, the frequency summation can be performed analytically at $T=0$ introducing a finite electronic
broadening $\gamma_e$, such that the corresponding $T=0$ Green's function reads $G_{a,b}(\mathbf k, \omega) = [\omega-E_{a,b}(\mathbf k)+i\gamma_e \text{sgn}(\omega)]^{-1}$. The remaining momentum integration is evaluated
numerically over the honeycomb-lattice Brillouin zone. 

We consider different sets of model parameters chosen to reproduce the
characteristic electronic regimes and energy scales of the materials of
interest. For NbSe$_2$ [Fig.\ 1(c)-(f)], we choose $t=3$ eV, $\Delta=0.3$ eV, and $\mu=0.5$ eV. The in-plane transverse optical phonon frequency is set to $\omega_p=0.03$ eV \cite{phonNbSe2}, the bare cavity mode to $\omega_{c,0}=0.029$ eV, and the electron-phonon coupling to $\lambda=0.1$ eV. The electronic broadening is set to $\gamma_e=0.025$ eV, in agreement with the experimental values \cite{basov2001}. Polariton formation remains suppressed irrespective of the number of layers $n_z$. 
For hBN [Fig.\ 2(a)-(d)], we use $t=3$ eV, $\Delta=2$ eV, and $\mu=0$ eV, with $\omega_p=0.2$ eV, $\omega_{c,0}=0.195$ eV, and $\lambda=0.3$ eV. The number of layers is set to $n_z=50$. If $n_z$ increases, the polariton splitting is enhanced, in agreement with the experimental findings \cite{BarraBurillo2021}. 
For gapped graphene ($n_z=1$), we retain the same parameters as for hBN, except for a reduced gap $\Delta=0.1$ eV \cite{Bistoni_2019} [Fig.\ 2(e)-(f)]. The electronic broadening $\gamma_e$ is kept fixed in the three cases.




\end{document}